\documentclass{aa}  
\usepackage{graphicx}
\usepackage{txfonts}
\usepackage[version=3]{mhchem}
\usepackage[dvipsnames]{xcolor}
\usepackage{lscape}
\usepackage{ulem}
\usepackage{bm}
\usepackage{booktabs}
\usepackage{hyperref}
\usepackage[figuresright]{rotating}
\usepackage{mathtools}
\usepackage{float}

\defcitealias{Sil2025}{Paper~I}

\newcommand{\micron}{\textmu m\xspace}

\begin{document}

   \title{Chemistry and ro-vibrational excitation of \ce{CH+} \\ in the planetary nebula NGC 7027}

   \author{M. Sil\inst{\ref{inst1},\ref{inst2},\ref{inst7}} \and A. Faure\inst{\ref{inst1}} \and H. Wiesemeyer\inst{\ref{inst3}} \and P. Hily-Blant\inst{\ref{inst1}} \and T. Gonz\'{a}lez-Lezana\inst{\ref{inst5}} \and J. Forer \inst{\ref{inst6}} \and J. Loreau\inst{\ref{inst4}} \and F. Lique\inst{\ref{inst2}}}

   \institute{Univ. Grenoble Alpes, CNRS, IPAG, 38000 Grenoble, France\\
              \email{milan.sil@gapp.nthu.edu.tw, milansil.astro@gmail.com, alexandre.faure@univ-grenoble-alpes.fr}
              \label{inst1}
         \and
             Univ Rennes, CNRS, IPR (Institut de Physique de Rennes) - UMR 6251, F-35000 Rennes, France
             \label{inst2}
         \and
             Institute of Astronomy, Department of Physics, National Tsing Hua University, Hsinchu, Taiwan
             \label{inst7}
          \and
              Max-Planck-Institut für Radioastronomie, Auf dem Hügel 69, D-53121 Bonn, Germany
              \label{inst3}
              \and
           Instituto de F\'{i}sica Fundamental, IFF-CSIC, Serrano 123, 28006 Madrid, Spain
             \label{inst5}
              \and
              Columbia Astrophysics Laboratory, Columbia University, New York, New York 10027, USA
              \label{inst6}
          \and
              KU Leuven, Department of Chemistry, B-3001 Leuven, Belgium
              \label{inst4}}

  \abstract 
{Small carbon hydride cations, such as {methylidynium} (\ce{CH+}), are important in the chemistry of the interstellar medium. They participate in a network of gas-phase reactions with a range of molecular and atomic species and lead to the formation of diverse hydrocarbon products that, in turn, act as precursors to more complex carbon-chain and organic molecules. }
{\ce{CH+} is known to be a reactive ion that is quickly destroyed by H, \ce{H2}, and free electrons, which makes its excitation challenging to model because chemical formation and destruction rates should be considered along with the usual radiative and inelastic rates when solving the statistical equilibrium equations. This so-called chemical pumping or excitation effect, already evidenced in the literature, is examined here with the first set of ab initio state-resolved ro-vibrational (reactive and inelastic) collision data to model the observed \ce{CH+} line intensities and better constrain the physico-chemical conditions of the environment.}
{Multiple rotational and ro-vibrational transitions of \ce{CH+} detected toward the planetary nebula NGC 7027 are analyzed in this work.
The chemical structure of \ce{CH+} in NGC~7027 was modeled with the photoionization code  \textsc{Cloudy} using updated formation and destruction reaction rate coefficients for \ce{CH+}. The electron temperature and atomic and molecular gas densities were modeled as a function of position within the nebula. The nonlocal thermodynamic equilibrium analysis of the observed \ce{CH+} emission lines was then performed with the 1D code \textsc{Cloudy} and single-zone code \texttt{RADEX} using an accurate and comprehensive set of spectroscopic and inelastic collisional data. In a second approach, chemical formation and destruction rate coefficients of \ce{CH+} were implemented in \texttt{RADEX}. This code was combined with a Markov chain Monte Carlo sampling (performed in the \texttt{RADEX}-parameter space) in order to extract the best-fit \ce{CH+} column density and physical conditions from the observed line fluxes.
}
{Our \textsc{Cloudy} model reproduces the observed \ce{CH+} line fluxes to within a factor of 1.3 on average, with a maximum deviation of a factor of 3. It also suggests that the rotational and ro-vibrational \ce{CH+} lines originate from physically distinct regions within NGC~7027 that differ mostly in kinetic temperature. Our \texttt{RADEX} {models show} that chemical pumping significantly enhances the population of all levels above $(\upsilon=0, J=1)$, with a strong increase in the ro-vibrational line intensities within the $\upsilon=2 \to 1$ band. A single-zone model, however, remains limited, and we strongly encourage using a full 1D model, which consistently incorporates all excitation processes, with the rate coefficients we accurately determined here.
}
   {}
   
   \maketitle
   \nolinenumbers
\section{Introduction \label{sec:intro}}

The source NGC~7027 is one of the brightest and best-studied young planetary nebulae. It is characterized by a dense, carbon-rich molecular envelope surrounding a hot central star \citep[hereafter, \citetalias{Sil2025} and references therein]{Sil2025}.
It serves as a classic example for examining the interactions between intense ultraviolet radiation and molecular chemistry in evolved stellar environments.
Despite the harsh ultraviolet radiation field, a rich molecular chemistry is sustained in its outer layers, where vibrationally excited \ce{H2} and \ce{CH+} have been observed in the photodissociation region (PDR) of NGC~7027.
Methylidynium (\ce{CH+}) was initially detected in the planetary nebula NGC~7027 more than two decades ago through far-infrared observations of its pure rotational transitions \citep{Cernicharo1997}.
\cite{neuf20} detected nine ro-vibrational {emission lines} of \ce{CH+}, $\upsilon=1\to0~R(0)-R(3)$ and $P(1)-P(5)$ for the first time in the same source.
Subsequently, \cite{neuf21} reported the detection of four additional ro-vibrational emission lines of {\ce{CH+} $\upsilon = 1\to0$, $P(7)-P(10)$ transitions} along with nine infrared transitions of \ce{H2}, comprising the $S(8), S(9), S(13)$, and $S(15)$ pure rotational lines, $\upsilon = 1\to0 \ O(4)-O(7)$ lines, and the $\upsilon = 2\to1 \ O(5)$ line.
The presence of \ce{CH+} in this environment is particularly significant because its efficient formation is attributed to the endothermic reaction C$^+(^2P)$ + H$_2(^1\Sigma^+) \to$ CH$^+(X^1\Sigma^+)$ + H$(^2S)$, which becomes exothermic when \ce{H2} is rotationally ($J \geq 8$, where $J$ is the rotational quantum number) or vibrationally ($\upsilon \geq 1$, where $\upsilon$ is the vibrational quantum number) excited, highlighting the key role of internally excited \ce{H2} in the chemistry within the nebula \citep{Godard2013,Faure2017,neuf21}.
If \ce{CH+} is destroyed before reaching a radiative or collisional equilibrium, the level populations will (partly) reflect its formation conditions. This effect is known as chemical formation pumping and was originally investigated in detail by \cite{Godard2013}.
Later, using the first accurate set of state-resolved rate coefficients for \ce{CH+} rotational excitation (with $J\leq 13$) \citep{Faure2017} and then a larger set with approximate data for \ce{CH+} ro-vibrational excitation \citep{neuf21}, it was confirmed that chemical pumping indeed has a substantial effect on the excitation of \ce{CH+} in PDRs. 

First identified in absorption in the diffuse interstellar medium (ISM) by \cite{Douglas1941}, \ce{CH+} has subsequently been observed in a variety of astrophysical environments beyond planetary nebulae, including massive star-forming regions \citep{Falgarone2010}, protoplanetary disks \citep{Thi2011,Zannese2025}, PDRs, such as the Orion bar \citep{Zannese2025}, and external galaxies \citep{Rangwala2014}.
We note, in particular, that through the unprecedented spatial resolution and sensitivity of the James Webb Space Telescope (JWST), \ce{CH+} and \ce{CH3+} ions have been detected in emission in the Orion bar and in d203-506 protoplanetary disk regions, heavily irradiated by the Trapezium star cluster \citep{Berne2023,Zannese2025}.

We report new calculations for the chemistry and excitation of \ce{CH+} in NGC~7027 based on an accurate set of state-resolved rate coefficients covering the rotational and ro-vibrational excitation of \ce{CH^+}. This was made possible through recent theoretical and experimental works providing state-resolved data for the formation, excitation, and destruction of \ce{CH+}, as presented in Appendices~\ref{sec:app_A}, \ref{sec:app_B}, and \ref{sec:app_C}.
The nonlocal thermodynamic equilibrium (NLTE) analysis of the observed \ce{CH+} rotational and ro-vibrational emissions is first performed with the code \textsc{Cloudy} \citep[v23.01;][]{chat23} and then with the code \texttt{RADEX} \citep{vand07}. Both codes use our newly constructed \ce{CH+} collision dataset for inelastic transitions.
In a second approach, chemical pumping of \ce{CH+} is implemented in \texttt{RADEX}, and the observed line fluxes are fitted using Markov chain Monte Carlo (MCMC) sampling \citep{Goodman2010} performed on the parameter space, namely the kinetic temperature, the atomic hydrogen and electron densities, and the \ce{CH+} column density.
In Sect.~\ref{sec:model} we report the predicted results of our NLTE models and compare them with the observed \ce{CH+} line fluxes.
Finally, in Sect.~\ref{sec:conclusions}, we briefly discuss the implications of our important findings and summarize our concluding remarks.


\section{Modeling and results} \label{sec:model}

\subsection{\textsc{Cloudy} modeling and results}
\label{sec:cloudy}

We used the 1D photoionization code \textsc{Cloudy} to model the chemical structure of \ce{CH+} and to interpret the observed line flux features in the prototypical PDR of NGC~7027, in a similar way as in \citetalias{Sil2025} (which focused on \ce{HeH+}). To do this, we approximated the elongated nebula to be spherically symmetric and assumed that the dynamically young nebula was in steady-state equilibrium.

\subsubsection{Physical conditions} \label{sec:phy}

The physical model parameters are listed in \autoref{tab:phy}. Our standard model assumes a homogeneous spherical shell of gas surrounding a hot central star, a white dwarf, emitting as a blackbody. This model is computed under the assumption of constant pressure, as in \citetalias{Sil2025}. The pressure was adjusted by fixing the initial total hydrogen nuclei density [$n({\rm H_{tot}})$] at $3.06\times10^4$~cm$^{-3}$ to obtain a Str\"{o}mgren sphere of angular radius at 4.6\arcsec ($\sim$ 0.02186~pc = $6.74\times10^{16}$~cm,
corresponding to the geometric mean of the semimajor and semiminor axes of the radio continuum emission; \citealp{basart87}),
considering a source distance of 980~pc \citep{zijl08}.
The resulting geometry is a thick, spherical, and closed shell.
The default initial gas-phase elemental abundance (relative to the total number of hydrogen nuclei) set for planetary nebulae in \textsc{Cloudy} modeling is primarily from \citet[][see Table~2 of \citetalias{Sil2025}]{alle83,khro89}.
We only highlight the modifications made to the modeling of NGC~7027 compared to \citetalias{Sil2025}, while more detailed modeling is presented in \citetalias{Sil2025}.

\begin{table}[h!]
\caption{Physical parameters and corresponding values used in the \textsc{Cloudy} model.}
	\label{tab:phy}
\centering
\resizebox{\linewidth}{!}{\begin{tabular}{lc}
\hline
\hline
{\bf Parameters} & {\bf Values}  \\
\hline
            Geometry & Closed thick shell \\
		Distance & 980 pc ($\sim 3.02\times10^{21}$ cm) \\
		Inner angular radius & 3.1\arcsec ($\sim$ 0.0147 pc = $4.55\times10^{16}$ cm) \\
		Str\"{o}mgren angular radius & 4.6\arcsec ($\sim$ 0.0218 pc = $6.75\times10^{16}$ cm) \\
            Stellar luminosity & $6.2\times10^3$~L$_{\sun} = 2.37\times10^{37}$~erg~s$^{-1}$ \\
            Stellar effective temperature & $1.98\times10^5$~K \\
            Radiation field & \cite{blac87} + \cite{Draine1996} + CMB \\
            Initial Hydrogen density [$n({\rm H_{tot}})$] & $3.06\times10^4$ cm$^{-3}$ \\
            Turbulence ($\sigma_{turb}$) & 6~km~s$^{-1}$\\
            Total pressure & Constant ($\sim 2\times10^9$~K~cm$^{-3}$) \\
            Mean primary CR ionization rate & $2\times10^{-16}$ s$^{-1}$ \\
            Type of grain & mix of amorphous carbon LG, VSG, and PAH \\
            Stopping A$_V$ & 1.50~mag \\
            Dust-to-gas mass ratio & $1/145$ \\
\hline
\end{tabular}}
\end{table}

Following \cite{Lau2016}, we adjusted the dust properties of three independent components with large grains (LGs), very small grains (VSGs), and polycyclic aromatic hydrocarbons (PAHs), constituting exactly 96.6\%, 1.9\%, and 1.5\% of the total dust by mass, respectively.
With these different components of the grain size distribution, we obtained a dust-to-gas mass ratio ($M_d/M_g$) of $\sim 1/145$.
A ratio of extinction per reddening of $R_V=A_V/E(B-V)=7.71$ and an extinction-to-gas ratio = $A_V/N(H) \sim 1.37\times10^{-22}$~mag~cm$^2$ were obtained from our model.
For comparison, typical diffuse ISM values are $R_V \approx 3.1$ and $A_V/N(H) \approx 5.3 \times 10^{-22}$~mag~cm$^2$. 
The significantly higher $R_V$ and lower $A_V/N(H)$ obtained from our model indicate a flatter extinction curve and reduced extinction efficiency per hydrogen atom relative to the diffuse ISM, consistent with a grain population biased toward larger sizes and the dust mass distribution being dominated by large grains, with a relative deficit of small grains.

As a caveat, we note that the sound travel time ($\sim 4186$~years) in our isobaric model is longer than the kinematic \citep[$\sim 600$~years;][]{mass89} and dynamic \citep[$\sim 1000-2000$~years;][]{Santander2012,Ali2015,scho18,guer20} ages of the nebula. 
The formation and destruction timescales ($\sim 10^4$~years) of \ce{H2} are also much longer than the age of the system, which makes the steady-state equilibrium assumption practical, but not fully justified.
The turbulent velocity dispersion ($\sigma_{turb}$) was set to 6~km~s$^{-1}$ following \cite{neuf21}.
Finally, our model incorporates the comprehensive model of the \ce{H2} molecule as described by \cite{shaw05}.
In the following, we present the results of our model.

\subsubsection{Chemistry} \label{sec:chemistry}

Given the physical conditions described above, the code computed the abundance of each species by solving the chemical reactions under steady-state assumptions.
The equilibrium abundance of \ce{CH+} is dominated by the reactions noted in \autoref{tab:updated_reaction} (see \autoref{sec:app_A}), specifically, the formation reaction between \ce{C+} and \ce{H2} (reaction R1) and destruction reaction with atomic hydrogen (reaction R4) due to their high densities in the PDR.
All the rate coefficients of the reactions noted in Table~\ref{tab:updated_reaction} were updated using recent calculations, except for reaction R5 (for more details, see \autoref{sec:app_A}).
The other chemical reactions related to \ce{CH+} followed the default settings in \textsc{Cloudy} v23.01.

\begin{figure}
    \includegraphics[width=0.5\textwidth]{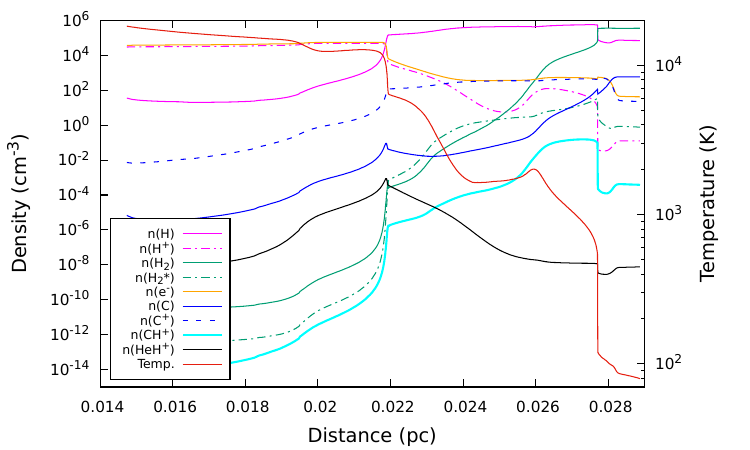}
    \includegraphics[width=0.5\textwidth]{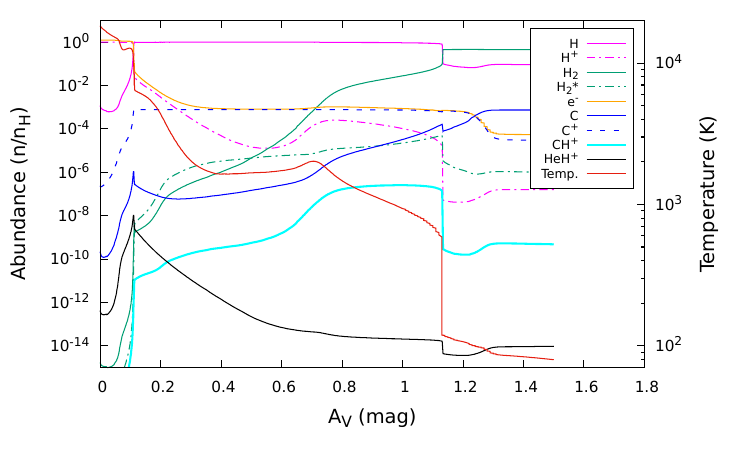}
    \includegraphics[width=0.5\textwidth]{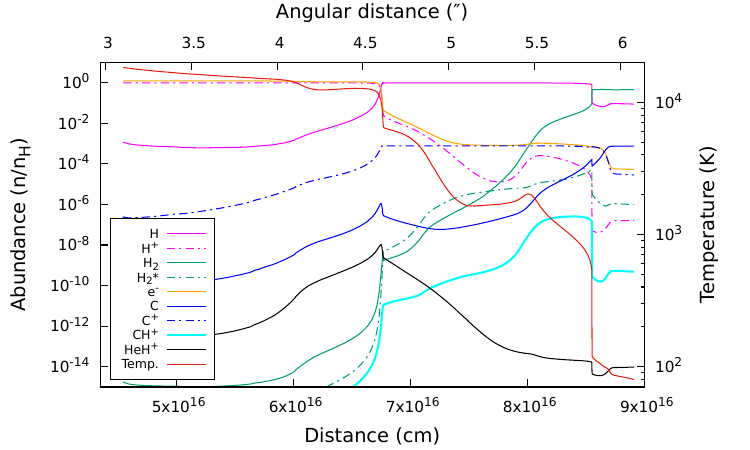}
    \caption{Top panel: Equilibrium temperature and densities of H, \ce{H+}, \ce{H2}, H$_2^*$, e$^-$, C, \ce{C+}, \ce{CH+}, and \ce{HeH+} obtained for an isobaric (total pressure $\sim 2.0\times10^9$~K~cm$^{-3}$) \textsc{Cloudy} model with an initial total hydrogen nuclei density $n({\rm H_{tot}}) = 3.06 \times 10^4$~cm$^{-3}$ as a function of the distance from the star starting from the illuminated face into the depth of the cloud.
    Middle panel: Equilibrium temperature and abundances of species as a function of visual extinction ($A_V$).
    Bottom panel: Equilibrium temperature and abundances of species as a function of the distance (linear and angular) from the star, starting from the illuminated face into the depth of the cloud.
    \label{fig:abundance_density_profile}}
\end{figure}

The top panel of \autoref{fig:abundance_density_profile} shows the equilibrium temperature and density profiles of H, \ce{H+}, \ce{H2}, H$_2^*$ (vibrationally excited \ce{H2}), e$^-$, C, \ce{C+}, \ce{CH+}, and \ce{HeH+} as a function of the distance from the star, starting from the illuminated face ($\sim 0.0147$~pc, where the starlight strikes the cloud) up to a certain depth ($A_V=1.5$~mag) where the temperature drops to 80~K (distance of the outer model boundary $\sim 0.0289$~pc).
In the middle and bottom panels, we present the equilibrium temperature and abundances of species as functions of $A_V$ and the distance (linear and angular) from the star (starting from the ionization front), respectively.
The middle and bottom panels are shown explicitly for comparison with the left panel of Fig.~9 in \cite{neuf21}, where
the authors used the 1D \texttt{Meudon PDR} code \citep{LePetit_2006} in a plane-parallel setup to model \ce{CH+} for an isochoric PDR model with a density $n(\rm{H_{tot}}) = 3 \times 10^5$~cm$^{-3}$.
We obtained a \ce{CH+} peak abundance (always with respect to the total number of hydrogen nuclei, unless otherwise stated) of $\sim 2.6\times10^{-7}$ along the \ce{CH+} plateau region ($\sim 0.0265-0.0275$~pc with temperatures of $\sim 800-1500$~K). This is lower by a factor of 5 than that obtained ($\sim 1.2\times10^{-6}$) by \cite{neuf21} around temperatures $\sim 1000-2000$~K for an isochoric model (see their Fig.~9).
It is important to note that the \ce{CH+} abundance peak from our model (which appears at an angular distance of 5.6\arcsec$-$5.8\arcsec from the central star) shifts toward the outer boundary of the shell as compared to \cite{neuf21}, who obtained it around 4.8\arcsec$-$4.9\arcsec (see their left panel of Fig.~9).
Thus, our model suggests that the emission of \ce{CH+} is likely to originate from regions approximately at and extending beyond a spherical shell radius of 5.5\arcsec, consistent with the observational angular sizes for pure rotational lines of \ce{CH+} (see notes in \autoref{tab:line_fluxes_cloudy}).
The \ce{CH+} abundance peak plateau falls rapidly at the \ce{H/H2} transition front (the so-called dissociation front where molecular hydrogen number density reaches 50\% of the number density of total hydrogen nuclei) around $A_V = 1.13$~mag.
We note that the \ce{H/H2} front is shifted outward (beyond $A_V = 1$~mag) compared to the findings of \cite{Godard2013,neuf21}.
We identified this effect as resulting from the modified (enhanced) rate coefficient for the destruction of \ce{H_2} by C$^+$ and from the dust model.

The \ce{HeH+} density and abundance profiles are also plotted in order to compare them with the findings from \citetalias{Sil2025}. As in our previous work, the \ce{HeH+} peak appears near the ionization front (where half of the \ce{H+} has recombined) with a peak abundance of $1.01\times10^{-8}$ and a total column density of $1.09\times10^{12}$~cm$^{-2}$, very similar to the results of \citetalias{Sil2025}.
This illustrates the benefit of using the code \textsc{Cloudy}, which is capable of describing the physical structure of H{\sc{ii}} regions (from the illuminated face of the cloud to the ionization front) and PDRs.

\subsubsection{NLTE calculations} \label{sec:NLTE_cloudy}

\begin{table*}
\caption{Comparison between the observed and \textsc{Cloudy} model-predicted line fluxes.}
	\label{tab:line_fluxes_cloudy}
\resizebox{\linewidth}{!}{\begin{tabular}{{lcccccc}}
 \hline
 \hline
 {\bf HeH$^+$ lines} & {\bf Rest wavelength} & {\bf Upper state} &  {\bf Observed line flux} & \multicolumn{2}{c}{\bf \textsc{Cloudy} model predicted} & {\bf Line flux ratio} \\
 $\bm{(\upsilon^\prime, J^\prime)\to(\upsilon, J)}$ & {\bf (\micron)} &  {\bf energy / $\bm{k_B}$ (K)} & {\bf ($\bm{10^{-18}}$~W~m$\bm{^{-2}}$)} & {\bf line flux($\dagger$)} & {\bf optical depth} & {\bf (Observed / Predicted)} \\
  & & & & {\bf ($\bm{10^{-18}}$~W~m$\bm{^{-2}}$)} & ($\bm{\tau}$) & $\bm{(F_{\rm obs}/F_{\rm pre})}$ \\
  \hline
  $(0, 1) \rightarrow (0,0)$ & 149.091 & 96 & $163\pm32^{(a)}$ & 49.14 & $4.92\times10^{-2}$ & $3.32 \pm 0.651$ \\
  $(1, 0) \rightarrow (0,1)\ [P(1)]$ & 3.51532 & 4188 & $1.55\pm0.16^{(b)}$  & 0.83 & $7.26\times10^{-5}$ & $1.87 \pm 0.193$  \\
  $(1, 1) \rightarrow (0,2)\ [P(2)]$ & 3.60677 & 4277 & $2.08\pm0.31^{(b)}$ & 0.93 & $4.50\times10^{-6}$ & $2.25 \pm 0.335$ \\
  \hline
  \hline
 {\bf Recombination lines} & {\bf Rest wavelength} & {\bf Upper state} &  {\bf Observed line flux$^{\bm{(b)}}$} & \multicolumn{2}{c}{\bf \textsc{Cloudy} model predicted} & {\bf Line flux ratio} \\
 & {\bf (\micron)} &  {\bf energy / $\bm{k_B}$ (K)} & {\bf ($\bm{10^{-18}}$~W~m$\bm{^{-2}}$)} & {\bf line flux($\dagger$)} & {\bf optical depth} & {\bf (Observed / Predicted)} \\
  & & & & {\bf ($\bm{10^{-18}}$~W~m$\bm{^{-2}}$)} & ($\bm{\tau}$) & $\bm{(F_{\rm obs}/F_{\rm pre})}$ \\
  \hline 
  H\,{\sc i} $19-6$ & 3.64493 & --- & $23.9\pm0.22$ & 20.52 & --- & $1.16 \pm 0.011$ \\
        He\,{\sc ii} $13-9$ & 3.54328 & --- & $53.7\pm0.21$ & 44.17 & --- & $1.22 \pm 0.005$ \\
        He\,{\sc i} $5^3D-4^3P^0$ & 3.70256 & --- & $8.44\pm0.48$ & 11.39 & --- & $0.74 \pm 0.042$ \\
  \hline
  \hline
{\bf \ce{CH+} rotational lines} & {\bf Rest} & {\bf Upper state} & {\bf Observed line flux} & \multicolumn{2}{c}{\bf \textsc{Cloudy} model predicted} & {\bf Line flux ratio} \\
 $\bm{(\upsilon^\prime, J^\prime)\to(\upsilon, J)}$  & {\bf frequency / wavelength} & {\bf energy / $\bm{k_B}$ (K)} & {\bf ($\bm{10^{-19}}$~W~cm$\bm{^{-2}}$)} &  {\bf line flux(*)} & {\bf optical depth} & {\bf (Predicted / Observed)} \\
& {\bf (GHz / \micron)} & & & {\bf ($\bm{10^{-19}}$~W~cm$\bm{^{-2}}$)} & ($\bm{\tau}$) & $\bm{(F_{\rm pre}/F_{\rm obs})}$ \\
\hline
$(0, 1) \rightarrow (0,0)$ & 835.1375 / 358.9738 & 40 & $0.47\pm0.01^{(c)}$ & 1.21 & $9.19$ & $2.57\pm0.055$ \\
$(0, 2) \rightarrow (0,1)$ & 1669.2820 / 179.5937 & 120 & $1.51\pm0.05^{(d)}$  & 4.54 & $10.69$ & $3.01\pm0.100$ \\
$(0, 3) \rightarrow (0,2)$ & 2501.4430 / 119.8478 & 240 & $2.18\pm0.17^{(d)}$  & 3.67 & 1.44 & $1.69\pm0.131$ \\
$(0, 4) \rightarrow (0,3)$ & 3330.6350 / 90.0106 & 400 & $2.00\pm0.22^{(d)}$  & 2.70 & $8.14\times10^{-2}$ & $1.35\pm0.149$ \\
$(0, 5) \rightarrow (0,4)$ & 4155.8795 / 72.1369 & 600 & $2.50\pm0.41^{(d)}$ & 1.88 & $9.86\times10^{-3}$ & $0.75\pm0.123$ \\
$(0, 6) \rightarrow (0,5)$ & 4976.2080 / 60.2452 & 838 & $2.41\pm0.33^{(d)}$ & 1.53 & $2.27\times10^{-3}$ & $0.63\pm0.087$ \\
\cline{7-7}
&&&&&& {$\exp{(\mu)}$ = 1.43$^{(\ddagger)}$} \\ 
\hline
\hline      
{\bf \ce{CH+} ro-vibrational lines} & {\bf Rest wavelength} & {\bf Upper state} &  {\bf Observed line flux$^{\bm{(e)}}$} & \multicolumn{2}{c}{\bf \textsc{Cloudy} model predicted} & {\bf Line flux ratio} \\
 $\bm{(\upsilon^\prime, J^\prime)\to(\upsilon, J)}$ & {\bf (\micron)} &  {\bf energy / $\bm{k_B}$ (K)} & {\bf ($\bm{10^{-18}}$~W~m$\bm{^{-2}}$)} & {\bf line flux(*)} & {\bf optical depth} & {\bf (Predicted / Observed)} \\
  & & & & {\bf ($\bm{10^{-18}}$~W~m$\bm{^{-2}}$)} & ($\bm{\tau}$) & $\bm{(F_{\rm pre}/F_{\rm obs})}$ \\
  \hline
	$(1, 1) \rightarrow (0,0)\ [R(0)]$ & 3.6146 & 3980 & $2.35\pm0.09$ & 2.44  & $1.35\times10^{-3}$ & $1.04\pm0.040$ \\
           $(1,2) \rightarrow (0,1)\ [R(1)]$ & 3.5811 & 4058 & $2.18\pm0.06$ & 2.87 & $7.67\times10^{-4}$ & $1.32\pm0.036$ \\
            $(1,3) \rightarrow (0,2)\ [R(2)]$ & 3.5496 & 4174 & $1.47\pm0.31$  & 1.59  & $6.80\times10^{-5}$ & $1.08\pm0.228$ \\
            $(1, 4) \rightarrow (0,3)\ [R(3)]$ & 3.5199 & 4328 & $1.09\pm0.06$  & 1.02 & $2.61\times10^{-6}$ & $0.93\pm0.051$ \\
            $(1, 0) \rightarrow (0,1)\ [P(1)]$ & 3.6876 & 3942 & $3.97\pm0.17$  & 6.01 & $8.83\times10^{-4}$ & $1.51\pm0.065$ \\
            $(1, 1) \rightarrow (0,2)\ [P(2)]$ & 3.7272 & 3980 & $7.38\pm0.11$ & 8.99 & $1.88\times10^{-4}$ & $1.22\pm0.018$ \\
            $(1, 2) \rightarrow (0,3)\ [P(3)]$ & 3.7689 & 4058 & $8.33\pm0.11$ & 12.21 & $1.57\times10^{-5}$ & $1.47\pm0.019$ \\
            $(1, 3) \rightarrow (0,4)\ [P(4)]$ & 3.8129 & 4174 & $7.67\pm0.15$ & 9.65 & $2.77\times10^{-6}$ & $1.26\pm0.025$ \\
            $(1, 4) \rightarrow (0,5)\ [P(5)]$ & 3.8591 & 4328 & $7.21\pm0.07$  & 9.96 & $9.06\times10^{-7}$ & $1.38\pm0.013$ \\
            $(1, 5) \rightarrow (0,6)\ [P(6)]$ & 3.9078  & 4520 & $6.45\pm0.21$  & 8.98 & $4.17\times10^{-7}$ & $1.39\pm0.045$ \\
            $(1, 6) \rightarrow (0,7)\ [P(7)]$ & 3.9589 & 4751 & $6.52\pm0.08$ & 7.87 & $2.02\times10^{-7}$ & $1.21\pm0.015$ \\
            $(1, 7) \rightarrow (0,8)\ [P(8)]$ & 4.0125 & 5019 & $5.04\pm0.09$  & 6.54 & $1.03\times10^{-7}$ & $1.30\pm0.023$ \\
            $(1, 8) \rightarrow (0,9)\ [P(9)]$ & 4.0688 & 5324 & $4.20\pm0.13$ & 4.78 & $5.57\times10^{-8}$ & $1.14\pm0.035$ \\
            $(1, 9) \rightarrow (0, 10)\ [P(10)]$ & 4.1278 & 5667 & $3.46\pm0.19$  & 3.43 & $3.12\times10^{-8}$ & $0.99\pm0.054$ \\
            \cline{7-7}
&&&&&& { $\exp{(\mu)}$ = 1.22$^{(\ddagger)}$} \\ 
\hline        
\hline      
{\bf H$_2$ lines} & {\bf Rest wavelength} & {\bf Upper state} &  {\bf Observed line flux$^{\bm{(e)}}$} & \multicolumn{2}{c}{\bf \textsc{Cloudy} model predicted} & {\bf Line flux ratio} \\
 $\bm{(\upsilon^\prime, J^\prime)\to(\upsilon, J)}$  & {\bf (\micron)} &  {\bf energy / $\bm{k_B}$ (K)} & {\bf ($\bm{10^{-18}}$~W~m$\bm{^{-2}}$)} & {\bf line flux(*)} & {\bf optical depth} & {\bf (Predicted / Observed)} \\
& & & & {\bf ($\bm{10^{-18}}$~W~m$\bm{^{-2}}$)} & ($\bm{\tau}$) & $\bm{(F_{\rm pre}/F_{\rm obs})}$ \\
  \hline
  $(0, 10) \rightarrow (0, 8)\ [S(8)]$ & 5.05174 & 8677 & $11.60\pm0.23$ & 71.16 & $3.52\times10^{-7}$ & $6.13\pm0.122$ \\
  $(0, 11) \rightarrow (0, 9)\ [S(9)]$ & 4.69333 & 10\,261 & $19.04\pm0.16$ & 125.38 & $4.37\times10^{-7}$ & $6.59\pm0.055$ \\
  $(0, 15) \rightarrow (0, 13)\ [S(13)]$ & 3.84506 & 17\,443 & $3.27\pm0.10$ & 15.28 & $1.33\times10^{-8}$ & $4.67\pm0.143$ \\
  $(0, 17) \rightarrow (0, 15)\ [S(15)]$ & 3.62518 & 21\,411 & $1.32\pm0.08$ & 12.48 & $2.18\times10^{-9}$ & $9.46\pm0.573$ \\
  $(1, 2) \rightarrow (0, 4)\ [O(4)]$ & 3.00305 & 6471 & $23.91\pm0.16$ & 134.35 & $1.06\times10^{-6}$ & $5.62\pm0.038$ \\
  $(1, 3) \rightarrow (0, 5)\ [O(5)]$ & 3.23411 & 6951 & $41.78\pm0.05$ & 256.00 & $1.60\times10^{-6}$ & $6.13\pm0.007$ \\
  $(1, 4) \rightarrow (0, 6)\ [O(6)]$ & 3.49985 & 7584 & $9.79\pm0.17$ & 43.98 & $2.28\times10^{-7}$ & $4.49\pm0.078$ \\
  $(1, 5) \rightarrow (0, 7)\ [O(7)]$ & 3.80638 & 8365 & $11.70\pm0.11$ & 58.24 & $2.53\times10^{-7}$ & $4.98\pm0.047$ \\
  $(2, 3) \rightarrow (1, 5)\ [O(5)]$ & 3.43693 & 12\,550 & $2.09\pm0.11$ & 9.08 & $1.35\times10^{-8}$ & $4.35\pm0.229$ \\
  \cline{7-7}
&&&&&& {$\exp{(\mu)}$ = 5.66$^{(\ddagger)}$} \\ 
  \hline
  \hline
\end{tabular}}
\tablefoot{$^{(a)}$ \cite{gust19}, $^{(b)}$ \cite{neuf20}, $^{(c)}$ \cite{Wesson2010}, $^{(d)}$ \cite{Cernicharo1997}, $^{(e)}$ \cite{neuf21}. \\
($\dagger$) See the footnote of Table 3 of \citetalias{Sil2025}. \\
(*) The line fluxes in \textsc{Cloudy} are given in units of erg~s$^{-1}$~cm$^{-2}$. 
We converted them into beam-integrated fluxes by scaling to the observed angular sizes for a direct comparison with the observations.
The conversion formulas we used are \\ the predicted flux (W~cm$^{-2}$) for \ce{CH+} $\upsilon=0$ lines = \textsc{Cloudy} flux (erg~s$^{-1}$~cm$^{-2}$) $\times 10^{-7} \times \frac{1}{4{\rm\pi}} \times {\rm\pi} \times (5.5 \times \frac{\rm\pi}{180 \times 3600})^2$ {[where ${\rm\pi} \times (5.5\arcsec)^2$ is the sky-plane area subtended by a spherical nebula equivalent to that of the elliptical nebula].} \\
The predicted flux (W~m$^{-2}$) for \ce{CH+} $\upsilon=1 \rightarrow 0$ and H$_2$ lines = \textsc{Cloudy} flux (erg~s$^{-1}$~cm$^{-2}$) $\times 10^{-3} \times \frac{1}{4{\rm\pi}} \times (0.375 \times 15) \times (\frac{\rm\pi}{180 \times 3600})^2$ [where $0.375\arcsec \times 15\arcsec$ is the area covered by the IRTF/iSHELL slit]. \\
$^{(\ddagger)}$ The quantity $\exp{(\mu)}$ indicates the geometric mean of the factor by which the model reproduces the observations, where $\mu$ is the mean logarithmic ratio of model-predicted and observed line fluxes.
}
\end{table*}

\begin{figure*}
\sidecaption
    \includegraphics[width=12cm]{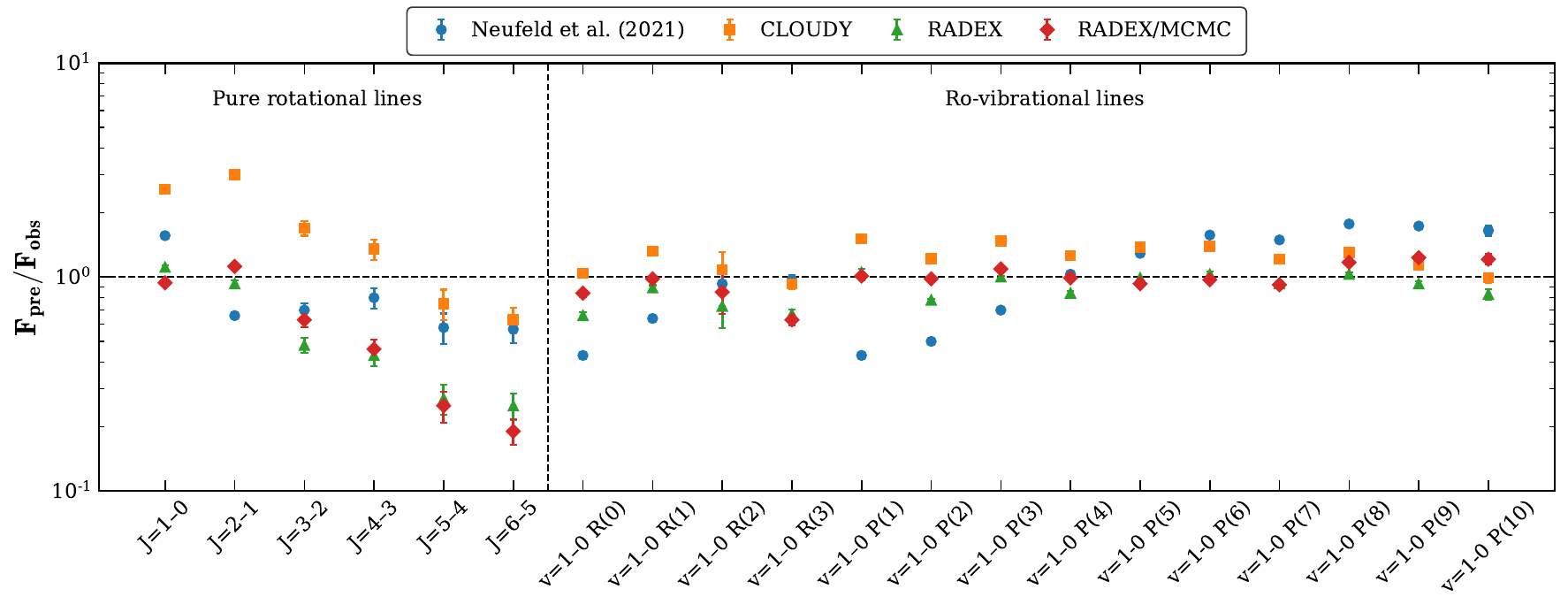}
    \caption{Ratios of different model-predicted to observed line fluxes ($F_{\rm pre}/F_{\rm obs}$) of rotational and ro-vibrational lines of \ce{CH+} as functions of the line names.
    \label{fig:CH+_line_ratio}}
\end{figure*}

\begin{figure}
    \includegraphics[width=0.49\textwidth]{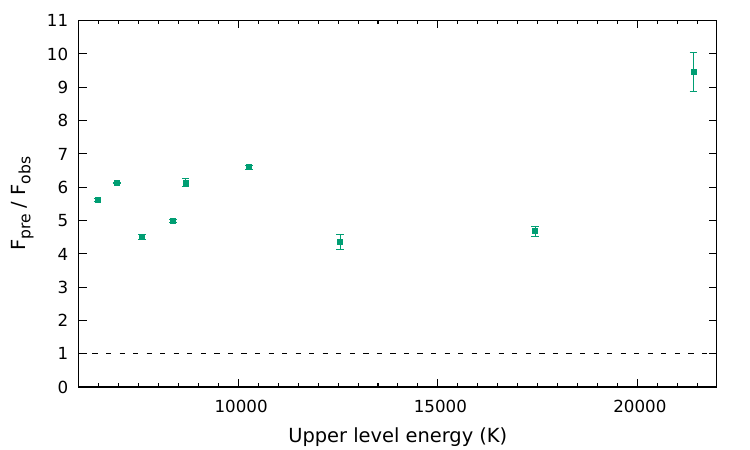}
    \caption{Ratios of \textsc{Cloudy} model-predicted to observed line fluxes ($F_{\rm pre}/F_{\rm obs}$) of rotational and ro-vibrational lines of \ce{H2} as a function of the upper level energies of transitions. 
    \label{fig:model_vs_obs_H2}}
\end{figure}

In \textsc{Cloudy}, the NLTE radiative transfer is treated in 1D geometry within the escape probability formalism \citep[][and references therein]{ferl17}. We employed the closed spherically symmetric geometry (implicit for planetary nebulae) with the \texttt{sphere expanding} (default) option so that the large velocity gradient (LVG), or Sobolev approximation, was employed. 

For the spectroscopy and collisional excitation of \ce{H_2}, the default \textsc{Cloudy} dataset was employed \citep{Wrathmall2007}. For \ce{CH+}, spectroscopic data (i.e., level energies and radiative rates) were extracted from the \texttt{EXOMOL} database \citep{tennyson2016,Pearce2024} and restricted to the first electronic state ($X^1\Sigma^+$). This set includes a total of 80 ro-vibrational levels, with the highest level ($\upsilon=3, J=11$) lying  9495~cm$^{-1}$ (13\,661~K) above the ground state, just below the first excited electronic state, $a^3\Pi$, at $\sim 1.2$~eV \citep{Amitay1996,Biglari2014}. For inelastic collisions of \ce{CH+} with hydrogen atoms and free electrons, we combined several accurate sets of theoretical data with approximations based on propensity rules in order to cover all 80 levels, as described in Appendix~\ref{sec:app_C}.

In \autoref{tab:line_fluxes_cloudy} we compare the observed line fluxes for \ce{CH+} and H$_2$ reported by \cite{neuf21}, along with those for 
\ce{HeH+} and H, \ce{He+} recombination lines reported by \cite{neuf20} to the predictions of our isobaric \textsc{Cloudy} model.
The line fluxes in \textsc{Cloudy} are given in units of erg~s$^{-1}$~cm$^{-2}$. 
We converted them into beam-integrated fluxes by scaling to the observed angular sizes (see notes in \autoref{tab:line_fluxes_cloudy}) for direct comparison with the observations.

As a goodness diagnostic, we report the mean logarithmic ratio of model-predicted ($F_{\rm pre}$) and observed ($F_{\rm obs}$) line fluxes,
\begin{equation}
\mu = \frac{1}{N} \sum_{i=1}^{N} \ln\left(\frac{F_{\rm pre}}{F_{\rm obs}}\right),
\end{equation}
where $N$ is the number of transitions.
The quantity $\exp{(\mu)}$ indicates the geometric mean of the factor by which the model reproduces the observations.

Overall, good agreement is obtained between our \textsc{Cloudy} model and observations (see orange squares in Fig.~\ref{fig:CH+_line_ratio}).
On average, the model reproduces observations to within a factor of $\exp{(\mu)}=1.28$.
To be more specific, our model overestimates the \ce{CH+} pure rotational lines from $J=1\rightarrow0$ to $J=4\rightarrow3$ by up to a factor of 3, while it underestimates the lines for $J=5\rightarrow4$ and $J=6\rightarrow5$ by factors of 1.3 and 1.6, respectively.
For the \ce{CH+} ro-vibrational lines, the agreement is within a factor of $\exp{(\mu)}=1.22$ on average.
A good match is thus obtained between our \textsc{Cloudy} model and observations, suggesting that our updated rate coefficients are reliable and that the chosen \ce{H2} excitation temperature (1900~K) for reaction R1 is appropriate for NGC~7027. 
It is instructive to compare our results with the best isobaric \texttt{Meudon PDR} model (including \ce{CH+} chemical pumping) of \cite{neuf21}, even though the reactive and inelastic rate coefficients were less accurate. Thus, with a thermal pressure of $P/k_B = 3\times10^8$~K~cm$^{-3}$, their model underestimates the pure rotational lines (except $J=1\to0$) and lower-energy ro-vibrational lines, while it overestimates high-excitation ro-vibrational lines of \ce{CH+} by up to a factor of 2. Thus, globally, the isobaric \textsc{Cloudy} and \texttt{Meudon PDR} models show comparable agreement with the observations (see blue and orange symbols in Fig.~\ref{fig:CH+_line_ratio}). 
We note that the constant-pressure prescription in \textsc{Cloudy} refers to total pressure (gas + turbulent + radiation), whereas in the \texttt{Meudon PDR} code, it corresponds to gas thermal pressure alone. Therefore, the numerical pressure values adopted in different codes are not directly equivalent.

In order to better highlight our results, 
Fig.~\ref{fig:emissivity_CH+} illustrates the intrinsic (local) line emissivities of \ce{CH+} as a function of distance from the star. The rotational and ro-vibrational \ce{CH+} emission peaks are seen to arise from two distinct physical regions that mainly differ in temperature ($\sim 850$ versus $1500$~K, respectively). We can also observe that the rotational emission peak shows a broader plateau, while the ro-vibrational peak is more sharply peaked. This result suggests that the pure rotational and ro-vibrational lines of \ce{CH+} probe separate regions in the PDR of NGC~7027, and this might help us to constrain the kinetic temperature gradient just upstream of the H/H$_2$ transition front.

A comparison between our \textsc{Cloudy} model-predicted and observed \ce{H2} line fluxes is listed in \autoref{tab:line_fluxes_cloudy} and depicted in Fig.~\ref{fig:model_vs_obs_H2}.
Our isobaric model is found to reasonably explain the observed fluxes of \ce{H2} lines to within a factor of 5.66 on average.
As a comparison, the best isobaric \texttt{Meudon PDR} model by \cite{neuf21} reproduces the \ce{H2} ro-vibrational line fluxes within a factor of 4, except for the $\upsilon=0\rightarrow0$ S(15) and S(13) lines of \ce{H2}, which are overestimated by a factor of $\sim 10$ and 20, respectively.
To facilitate the understanding of the emitting region of the observed \ce{H2} lines, the intrinsic \ce{H_2} line emissivities obtained from our \textsc{Cloudy} model are also plotted in Fig.~\ref{fig:emissivity_H2}.
Figs.~\ref{fig:emissivity_CH+} and \ref{fig:emissivity_H2} both show that \ce{CH+} emits from the same region as \ce{H2}, meaning that \ce{CH+} emission is strongly linked to vibrationally excited \ce{H2}, as expected. 

It should be noted that the dynamical age of NGC~7027 depends on the structures to which it refers.
For the inner region traced by the emissions modeled here, it is $1000-2000$~years \citep{Santander2012,Ali2015,scho18,guer20}, while earlier estimates based on multi-epoch observations of the nebula radio emission yield $\sim 600$~years \citep{mass89}.
In the PDR modeled here, \ce{H2} formation is primarily driven by grain-surface reactions and not by the \ce{H-} channel, even in the warm molecular layer where \ce{CH+} and \ce{H2} emissions arise at temperatures of $\sim1500$~K.
The discrepancy may be due to the time-dependent thermal and chemical evolution of the warm molecular gas, which cannot be captured by the current steady-state approximation.
This might explain why our stationary \textsc{Cloudy} model overpredicts the \ce{H2} emission.

 \autoref{tab:line_fluxes_cloudy} shows that for \ce{HeH+}, H~{\sc{i}}, He~{\sc{ii}}, and He~{\sc{i}}, our predictions are fully consistent with \citetalias{Sil2025}. This confirms that our current isobaric model is able to explain most of the observed line fluxes of \ce{CH+}, \ce{HeH+}, H$_2$, and H, \ce{He+} recombination lines simultaneously and within a factor of 2 to 5, with the exception of some ro-vibrational lines of \ce{H_2}.

\begin{table}
\caption{Physical parameters at \ce{CH+} abundance peak and the total \ce{CH+} column density obtained from the \textsc{Cloudy} model.}
	\label{tab:phy_ch+_peak}
\centering
\resizebox{\linewidth}{!}{\begin{tabular}{lcc}
\hline
\hline
{\bf Parameters} & \multicolumn{2}{c}{\bf Peak values at region of \ce{CH+}}  \\
& {\bf Rotational emission} & {\bf Ro-vibrational emission} \\
\hline
            Distance from the star & $\sim 0.027$ pc & $\sim 0.026$ pc \\ 
            Kinetic gas temperature, $T_{\rm kin}$ & 854~K & 1507~K \\
		Electron density, $n(\rm e^-)$ & $5.26\times10^2$~cm$^{-3}$ & $5.55\times10^2$~cm$^{-3}$\\
        H atom density, $n({\rm H})$ & $5.97\times10^5$~cm$^{-3}$ & $5.16\times10^5$~cm$^{-3}$ \\
        \ce{H2} density, $n(\ce{H2})$ & $2.46\times10^4$~cm$^{-3}$ & $3.73\times10^3$~cm$^{-3}$ \\
        \hline
  \ce{CH+} total column density, $N({\rm CH^+})$ & \multicolumn{2}{c}{$6.09\times10^{14}$~cm$^{-2}$} \\
\hline
\end{tabular}}
\end{table}

\begin{table*}
\caption{Comparison between the observed and \texttt{RADEX} model-predicted \ce{CH+} line fluxes obtained considering the physical parameters at \ce{CH+} ro-vibrational emission peak (noted in \autoref{tab:phy_ch+_peak}) from the \textsc{Cloudy} model$^a$.}
	\label{tab:line_fluxes_model_C}
\resizebox{\linewidth}{!}{\begin{tabular}{{lcccccc}}
 \hline
\hline
		{\bf \ce{CH+} rotational lines} & {\bf Rest frequency / wavelength} & {\bf Upper state} & {\bf Observed line flux} & \multicolumn{2}{c}{\bf \texttt{RADEX} model predicted} & {\bf Line flux ratio} \\
$\bm{(\upsilon^\prime, J^\prime)\to(\upsilon, J)}$  & {\bf (GHz / \micron)} & {\bf energy / $\bm{k_B}$ (K)} & {\bf ($\bm{10^{-19}}$~W~cm$\bm{^{-2}}$)} & {\bf line flux*} & {\bf Optical depth} & {\bf (Predicted / Observed)} \\
& & & & {\bf ($\bm{10^{-19}}$~W~cm$\bm{^{-2}}$)} & ($\bm{\tau}$) & $\bm{(F_{\rm pre}/F_{\rm obs})}$ \\
\hline
$(0, 1) \rightarrow (0,0)$ & 835.1375 / 358.9738 & 40 & $0.47\pm0.01$ & 0.52 & 2.77 & $1.11\pm0.024$ \\
$(0, 2) \rightarrow (0,1)$ & 1669.2820 / 179.5937 & 120 & $1.51\pm0.05$  & 1.41 & 1.72 & $0.93\pm0.031$ \\
$(0, 3) \rightarrow (0,2)$ & 2501.4430 / 119.8478 & 240 & $2.18\pm0.17$  & 1.05 & $1.63\times10^{-1}$ & $0.48\pm0.038$ \\
$(0, 4) \rightarrow (0,3)$ & 3330.6350 / 90.0106 & 400 & $2.00\pm0.22$  & 0.87 & $9.60\times10^{-3}$ & $0.43\pm0.048$ \\
$(0, 5) \rightarrow (0,4)$ & 4155.8795 / 72.1369 & 600 & $2.50\pm0.41$ & 0.67 & $1.90\times10^{-3}$ & $0.27\pm0.044$ \\
$(0, 6) \rightarrow (0,5)$ & 4976.2080 / 60.2452 & 838 & $2.41\pm0.33$ & 0.61 & $4.91\times10^{-4}$ & $0.25\pm0.035$ \\
\cline{7-7}
&&&&&& { $\exp{(\mu)}$ = 0.50$^{(\ddagger)}$} \\ 
\hline
\hline      
{\bf \ce{CH+} ro-vibrational lines} & {\bf Rest wavelength} & {\bf Upper state} &  {\bf Observed line flux} & \multicolumn{2}{c}{\bf \texttt{RADEX} model predicted} & {\bf Line flux ratio} \\
  $\bm{(\upsilon^\prime, J^\prime)\to(\upsilon, J)}$ & {\bf (\micron)} &  {\bf energy / $\bm{k_B}$ (K)} & {\bf ($\bm{10^{-18}}$~W~m$\bm{^{-2}}$)} & {\bf line flux*} & {\bf Optical depth} & {\bf (Predicted / Observed)} \\
  & & & & {\bf ($\bm{10^{-18}}$~W~m$\bm{^{-2}}$)} & ($\bm{\tau}$) & $\bm{(F_{\rm pre}/F_{\rm obs})}$ \\
  \hline
	$(1, 1) \rightarrow (0,0)\ [R(0)]$ & 3.6146 & 3980 & $2.35\pm0.09$ &  1.56 & $3.32\times10^{-4}$ & $0.66\pm0.025$ \\
           $(1,2) \rightarrow (0,1)\ [R(1)]$ & 3.5811 & 4058 & $2.18\pm0.06$ & 1.95 & $1.20\times10^{-4}$ & $0.89\pm0.025$ \\
            $(1,3) \rightarrow (0,2)\ [R(2)]$ & 3.5496 & 4174 & $1.47\pm0.31$  & 1.07 & $7.75\times10^{-6}$ & $0.73\pm0.153$ \\
            $(1, 4) \rightarrow (0,3)\ [R(3)]$ & 3.5199 & 4328 & $1.09\pm0.06$  & 0.73 & $3.32\times10^{-7}$ & $0.67\pm0.037$ \\
            $(1, 0) \rightarrow (0,1)\ [P(1)]$ & 3.6876 & 3942 & $3.97\pm0.17$  & 4.16 & $1.38\times10^{-4}$ & $1.05\pm0.045$ \\
            $(1, 1) \rightarrow (0,2)\ [P(2)]$ & 3.7272 & 3980 & $7.38\pm0.11$   & 5.77 & $2.15\times10^{-5}$ & $0.78\pm0.012$ \\
            $(1, 2) \rightarrow (0,3)\ [P(3)]$ & 3.7689 & 4058 & $8.33\pm0.11$  & 8.30 & $1.99\times10^{-6}$ & $1.00\pm0.013$ \\
            $(1, 3) \rightarrow (0,4)\ [P(4)]$ & 3.8129 & 4174 & $7.67\pm0.15$ & 6.48 & $5.56\times10^{-7}$ & $0.84\pm0.017$ \\
            $(1, 4) \rightarrow (0,5)\ [P(5)]$ & 3.8591 & 4328 & $7.21\pm0.07$  & 7.12 & $2.07\times10^{-7}$ & $0.99\pm0.01$ \\
            $(1, 5) \rightarrow (0,6)\ [P(6)]$ & 3.9078  & 4520 & $6.45\pm0.21$   & 6.63 & $1.06\times10^{-7}$ & $1.03\pm0.033$ \\
            $(1, 6) \rightarrow (0,7)\ [P(7)]$ & 3.9589 & 4751 & $6.52\pm0.08$   & 6.07 & $5.75\times10^{-8}$ & $0.93\pm0.011$ \\
            $(1, 7) \rightarrow (0,8)\ [P(8)]$ & 4.0125 & 5019 & $5.04\pm0.09$   & 5.20 & $3.26\times10^{-8}$ & $1.03\pm0.018$ \\
            $(1, 8) \rightarrow (0,9)\ [P(9)]$ & 4.0688 & 5324 & $4.20\pm0.13$  & 3.89 & $1.92\times10^{-8}$ & $0.93\pm0.029$ \\
            $(1, 9) \rightarrow (0,10)\ [P(10)]$ & 4.1278 & 5667 & $3.46\pm0.19$  & 2.87 & $1.17\times10^{-8}$ & $0.83\pm0.046$ \\      
            \cline{7-7}
&&&&&& $\exp{(\mu)}$ = 0.87$^{(\ddagger)}$ \\ 
\hline
\hline
\end{tabular}}
\tablefoot{
* \texttt{RADEX} line flux is given erg~s$^{-1}$~cm$^{-2}$.
We converted them into beam-integrated fluxes by scaling to the observed angular sizes (see notes in \autoref{tab:line_fluxes_cloudy}) for direct comparison with the observations. \\
$^a$ \textsc{Cloudy} parameters obtained at \ce{CH+} ro-vibrational emission peak: $T_{\rm kin}=1507$~K, $n({\rm e}^-)=5.55\times10^2$~cm$^{-3}$, $n({\rm H})=5.16\times10^5$~cm$^{-3}$, $N[{\rm CH^+}]=6.09\times10^{14}$~cm$^{-2}$, line width (FWHM) $=30$~km~s$^{-1}$. \\
$^{(\ddagger)}$ The quantity $\exp{(\mu)}$ indicates the geometric mean of the factor by which the model reproduces the observations, where $\mu$ is the mean logarithmic ratio of model-predicted and observed line fluxes.
}
\end{table*}

\subsection{\texttt{RADEX} modeling and results} \label{sec:radex}

In \texttt{RADEX}, the NLTE radiative transfer is solved for a homogeneous and isothermal medium within the escape probability formalism \citep{vand07}. As with \textsc{Cloudy}, the LVG approximation (option \texttt{expanding sphere}) was used to model the \ce{CH+} emission in NGC~7027.

The \texttt{RADEX} parameters for the physical conditions, namely, kinetic temperature $T_{\rm kin}$, atomic hydrogen density $n$(H), and electron density $n$(e$^-$), were extracted from the \textsc{Cloudy} solution at the \ce{CH+} abundance peak.
As previously discussed in Sect.~\ref{sec:chemistry} and illustrated in Fig.~\ref{fig:abundance_density_profile}, the \ce{CH+} abundance peak exhibits a plateau. 
Fig.~\ref{fig:emissivity_CH+} shows that the rotational and ro-vibrational \ce{CH+} lines originate from two distinct physical regions. As noted in Table~\ref{tab:phy_ch+_peak}, the main difference between these two regions is the temperature, which increases from 850~K (at the rotational emission peak) to 1500~K (at the ro-vibrational emission peak).
Since \texttt{RADEX} considers radiative transfer through a single-zone isothermal medium, radial 1D effects (e.g., varying physical and chemical conditions) can obviously not be captured. However, since the first excited vibrational state ($\upsilon=1$) of \ce{CH^+} opens at 3942~K above the ground state, it seems more appropriate to use a kinetic temperature above 1000~K in order to efficiently populate by collisions the observed \ce{CH+} ro-vibrational states.
We therefore selected the warmer physical conditions ($T=1507$~K) at the ro-vibrational emission peak, allowing the first vibrational state to be more easily populated through collisions via the high-energy tail of the Maxwellian kinetic energy distribution.
The last parameter is the \ce{CH+} column density per unit velocity interval. The total \ce{CH+} column density was taken as the \textsc{Cloudy} value (also reported in \autoref{tab:phy_ch+_peak}), and a line width (FWHM) of 30~km~s$^{-1}$ was selected as representative of the detected lines of \ce{CH+} \citep{Cernicharo1997,neuf21}.
The incident radiative continuum at the illuminated face of the cloud, derived from the \textsc{Cloudy} model {\citep[and in close agreement with the best-fit dust model of][]{Lau2016}}, was also included in \texttt{RADEX} as a background radiation field. We note, however, that the effect of this radiation field on the level populations is a few percent at most, so that radiative pumping of \ce{CH+} is minor in NGC~7027 (see discussion in Sect.~\ref{sec:excitation_comparison}).
Finally, the same spectroscopic and collisional dataset as used in \textsc{Cloudy} (see Sect.~\ref{sec:NLTE_cloudy}) was used, thus considering inelastic collisions with hydrogen atoms and electrons.

The predicted line fluxes are reported in \autoref{tab:line_fluxes_model_C}. The agreement is better for the ro-vibrational lines ($\exp{(\mu)}=0.87$) than for the rotational lines ($\exp{(\mu)}=0.50$), yielding an overall mean agreement of $\exp{(\mu)}=0.74$. 
Significant differences between \textsc{Cloudy} and \texttt{RADEX} are found for \ce{CH+} pure rotational lines, by factors of $\sim 2-3$, whereas for ro-vibrational lines, the \texttt{RADEX} line fluxes are only $\sim30\%$ lower than those from \textsc{Cloudy} (see orange and green symbols in Fig.~\ref{fig:CH+_line_ratio}, Tables~\ref{tab:line_fluxes_cloudy} and \ref{tab:line_fluxes_model_C}). The discrepancy observed for rotational lines reflects the extended plateau of rotational emissivities primarily, as illustrated in Fig.~\ref{fig:emissivity_CH+}.
Our \textsc{Cloudy} model indeed predicts a large gradient of the kinetic temperature across this region (just before the H/H$_2$ front). The inability of \texttt{RADEX} to simultaneously reproduce the observed \ce{CH+} rotational and ro-vibrational lines with comparable accuracy thus supports the finding that these lines probe distinct regions with different physical conditions. This is corroborated by the much better average factor ($\exp{(\mu)}$) for ro-vibrational lines compared to rotational lines. Finally, we note that the relatively good agreement between \textsc{Cloudy} and \texttt{RADEX} for the \ce{CH+} ro-vibrational line fluxes is similar to that observed in \citetalias{Sil2025} for \ce{HeH+}, whose abundance and line emissivities are even more strongly peaked.

\begin{table}
\caption{Summary of the initial priors and best-fit posterior physical parameters obtained from the \texttt{RADEX/MCMC} model (taking into account all rotational and ro-vibrational detected \ce{CH+} lines) with chemical pumping (see Fig.~\ref{fig:corner_plot_radmcmc_all}).}
	\label{tab:posterior}
\centering
\resizebox{\linewidth}{!}{\begin{tabular}{ccccc}
\hline
\hline
{\bf Parameters} & {\bf Units} & {\bf Initial priors} & {\bf Prior intervals} &  {\bf Best-fit posteriors}  \\
\hline
            & & & & \\
            $T_\mathrm{kin}$ & K & 1250 & $30-2000$ & $1258.61^{+11.40}_{-11.22}$ \\
            & & & & \\
            $n\mathrm{(e^-)}$ & cm$^{-3}$ & $1.00\times10^2$ & $10-10^5$
            & $5.89^{+0.28}_{-0.27}\times10^3$ \\
            & & & & \\
            $n({\rm H})$ & cm$^{-3}$ & $1.00\times10^6$ & $10^3-10^7$ & $1.62^{+0.16}_{-0.17}\times10^5$ \\
            & & & & \\
		  $N[\rm{CH^+}]$ & cm$^{-2}$ & $10^{15}$ & $10^{13}-10^{16}$ & $1.48^{+0.03}_{-0.03}\times10^{14}$ \\
  & & & & \\
\hline
\hline
\end{tabular}}
\tablefoot{
The uncertainties quoted represent the 16th and 84th percentiles (equivalent to a $1\sigma$ for a Gaussian distribution).
}
\end{table}

\begin{table*}
\caption{Comparison between the observed and the \texttt{RADEX/MCMC} best-fit model parameters$^a$ predicted \ce{CH+} line fluxes (taking into account all the rotational and ro-vibrational detected \ce{CH+} lines) with chemical pumping (see Fig.~\ref{fig:corner_plot_radmcmc_all}).}
	\label{tab:line_fluxes_model_mcmc}
\resizebox{\linewidth}{!}{\begin{tabular}{{lcccccc}}
 \hline
 \hline
		{\bf \ce{CH+} pure rotational lines} & {\bf Rest frequency / wavelength} & {\bf Upper state} & {\bf Observed line flux} & \multicolumn{2}{c}{\bf \texttt{RADEX} model predicted} & {\bf Line flux ratio} \\
$\bm{(\upsilon^\prime, J^\prime)\to(\upsilon, J)}$  & {\bf (GHz / \micron)} & {\bf energy / $\bm{k_B}$ (K)} & {\bf ($\bm{10^{-19}}$~W~cm$\bm{^{-2}}$)} & {\bf line flux*} & {\bf Optical depth} & {\bf (Predicted / Observed)} \\
& & & & {\bf ($\bm{10^{-19}}$~W~cm$\bm{^{-2}}$)} & ($\bm{\tau}$) &  $\bm{(F_{\rm pre}/F_{\rm obs})}$ \\
\hline
$(0, 1) \rightarrow (0,0)$ & 835.1375 / 358.9738 & 40 & $0.47\pm0.01$ & 0.44 & $2.40\times10^{-1}$ & $0.94\pm0.020$ \\
$(0, 2) \rightarrow (0,1)$ & 1669.2820 / 179.5937 & 120 & $1.51\pm0.05$  & 1.69 & $5.03\times10^{-1}$ & $1.12\pm0.037$ \\
$(0, 3) \rightarrow (0,2)$ & 2501.4430 / 119.8478 & 240 & $2.18\pm0.17$  & 1.38 & $1.10\times10^{-1}$ & $0.63\pm0.049$ \\
$(0, 4) \rightarrow (0,3)$ & 3330.6350 / 90.0106 & 400 & $2.00\pm0.22$  & 0.92 & $1.24\times10^{-2}$ & $0.46\pm0.050$ \\
$(0, 5) \rightarrow (0,4)$ & 4155.8795 / 72.1369 & 600 & $2.50\pm0.41$ & 0.63 & $2.09\times10^{-3}$ & $0.25\pm0.041$ \\
$(0, 6) \rightarrow (0,5)$ & 4976.2080 / 60.2452 & 838 & $2.41\pm0.33$ & 0.47 & $5.12\times10^{-4}$ & $0.19\pm0.026$ \\
\cline{7-7}
&&&&&& { $\exp{(\mu)}$ = 0.50$^{(\ddagger)}$} \\ 
\hline
\hline      
{\bf \ce{CH+} ro-vibrational lines} & {\bf Rest wavelength} & {\bf Upper state} &  {\bf Observed line flux} & \multicolumn{2}{c}{\bf \texttt{RADEX} model predicted} & {\bf Line flux ratio} \\
 $\bm{(\upsilon^\prime, J^\prime)\to(\upsilon, J)}$ & {\bf (\micron)} &  {\bf energy / $\bm{k_B}$ (K)} & {\bf ($\bm{10^{-18}}$~W~m$\bm{^{-2}}$)} & {\bf line flux*} & {\bf Optical depth} & {\bf (Predicted / Observed)} \\
  & & & & {\bf ($\bm{10^{-18}}$~W~m$\bm{^{-2}}$)} & ($\bm{\tau}$) &  $\bm{(F_{\rm pre}/F_{\rm obs})}$ \\
  \hline
	$(1, 1) \rightarrow (0,0)\ [R(0)]$ & 3.6146 & 3980 & $2.35\pm0.09$ &  1.97  & $4.79\times10^{-5}$ & $0.84\pm0.032$ \\
           $(1,2) \rightarrow (0,1)\ [R(1)]$ & 3.5811 & 4058 & $2.18\pm0.06$ & 2.13 & $3.83\times10^{-5}$ & $0.98\pm0.027$ \\
            $(1,3) \rightarrow (0,2)\ [R(2)]$ & 3.5496 & 4174 & $1.47\pm0.31$  & 1.25 & $5.45\times10^{-6}$ & $0.85\pm0.179$ \\
            $(1, 4) \rightarrow (0,3)\ [R(3)]$ & 3.5199 & 4328 & $1.09\pm0.06$  & 0.69 & $4.15\times10^{-7}$ & $0.63\pm0.035$ \\
            $(1, 0) \rightarrow (0,1)\ [P(1)]$ & 3.6876 & 3942 & $3.97\pm0.17$  & 3.99 & $4.40\times10^{-5}$ & $1.01\pm0.043$ \\
            $(1, 1) \rightarrow (0,2)\ [P(2)]$ & 3.7272 & 3980 & $7.38\pm0.11$ & 7.26 & $1.51\times10^{-5}$ & $0.98\pm0.015$ \\
            $(1, 2) \rightarrow (0,3)\ [P(3)]$ & 3.7689 & 4058 & $8.33\pm0.11$  & 9.06 & $2.49\times10^{-6}$ & $1.09\pm0.014$ \\
            $(1, 3) \rightarrow (0,4)\ [P(4)]$ & 3.8129 & 4174 & $7.67\pm0.15$  & 7.58 & $5.88\times10^{-7}$ & $0.99\pm0.019$ \\
            $(1, 4) \rightarrow (0,5)\ [P(5)]$ & 3.8591 & 4328 & $7.21\pm0.07$   & 6.73 & $1.95\times10^{-7}$ & $0.93\pm0.009$ \\
            $(1, 5) \rightarrow (0,6)\ [P(6)]$ & 3.9078  & 4520 & $6.45\pm0.21$   & 6.27 & $8.07\times10^{-8}$ & $0.97\pm0.032$ \\
            $(1, 6) \rightarrow (0,7)\ [P(7)]$ & 3.9589 & 4751 & $6.52\pm0.08$   & 5.97 & $4.45\times10^{-8}$ & $0.92\pm0.011$ \\
            $(1, 7) \rightarrow (0,8)\ [P(8)]$ & 4.0125 & 5019 & $5.04\pm0.09$   & 5.90 & $2.71\times10^{-8}$ & $1.17\pm0.021$ \\
            $(1, 8) \rightarrow (0,9)\ [P(9)]$ & 4.0688 & 5324 & $4.20\pm0.13$   & 5.17 & $1.68\times10^{-8}$ & $1.23\pm0.038$ \\
            $(1, 9) \rightarrow (0,10)\ [P(10)]$ & 4.1278 & 5667 & $3.46\pm0.19$  & 4.20 & $1.05\times10^{-8}$ & $1.21\pm0.067$ \\   
            \cline{7-7}
&&&&&& { $\exp{(\mu)}$ = 0.97$^{(\ddagger)}$} \\ 
\hline
\hline        
\end{tabular}}
\tablefoot{* \texttt{RADEX} line flux is given erg~s$^{-1}$~cm$^{-2}$. 
We converted them into beam-integrated fluxes by scaling to the observed angular sizes (see notes in \autoref{tab:line_fluxes_cloudy}) for direct comparison with the observations. \\
$^a$ \texttt{RADEX/MCMC} best-fitted model parameters from \autoref{tab:posterior}: $T_{\rm kin}=1258$~K, $n(\rm{e}^{-})=5.89\times10^3$~cm$^{-3}$, $n(\rm H)=1.62\times10^5$~cm$^{-3}$, $N[\ce{CH+}]=1.48\times10^{14}$~cm$^{-2}$, line width (FWHM) $=30$~km~s$^{-1}$. \\
$^{(\ddagger)}$ The quantity $\exp{(\mu)}$ indicates the geometric mean of the factor by which the model reproduces the observations, where $\mu$ is the mean logarithmic ratio of model-predicted and observed line fluxes.
}
\end{table*}

\subsection{\texttt{RADEX/MCMC} modeling and results} \label{sec:mcmc}

In this section, another approach that introduces chemical pumping into the NLTE analysis was used to investigate in depth the \ce{CH+} line excitation. Indeed, because \ce{CH+} belongs to the class of reactive ions, that is, species that react fast with the dominant colliders (here electrons and H atoms), the chemical source and sink terms must in principle be included when solving the statistical-equilibrium equations. These two terms describe the so-called chemical (or formation) pumping process by which a reactive ion (i.e., with short chemical lifetime) retains memory of its formation conditions. The general consequence is a hotter distribution of rotational and ro-vibrational levels.

The version of \texttt{RADEX}, as modified by \cite{Faure2017}, was employed in order to include the state-resolved formation and destruction rate coefficients of \ce{CH+} for all 80 \ce{CH+} levels belonging to the ground electronic state $X^1\Sigma^+$ (see Appendix~\ref{sec:app_B}).

In contrast to the previous (see Sect.~\ref{sec:radex}) NLTE \texttt{RADEX} calculations, where the physical conditions and \ce{CH+} total column density were taken from the \textsc{Cloudy} results (at \ce{CH+} ro-vibrational emission peak), \texttt{RADEX} was combined with MCMC sampling in order to determine without a model prescription the physical parameters that best reproduce all 20 lines of \ce{CH+} detected in NGC~7027.
MCMC is a statistical method used to estimate the parameters of a model by iteratively generating random samples from prior probability distributions. 
To compute the likelihood of the parameters given the observations, we used the publicly available MCMC Python implementation \texttt{emcee}\footnote{\url{https://emcee.readthedocs.io/en/stable/}} developed by \cite{foreman2013}.

As discussed above, the modified version of \texttt{RADEX} was used with the LVG approximation (option \texttt{expanding sphere}), a line width (FWHM) of 30~km~s$^{-1}$, and the background radiation field extracted from \textsc{Cloudy}.
We employed 32 MCMC walkers, each with 4096 steps, to ensure parameter space exploration and convergence.
We set the initial priors (noted in \autoref{tab:posterior}) based on the result obtained from the \textsc{Cloudy} model near the abundance peak of \ce{CH+} (see \autoref{tab:phy_ch+_peak}).
The corresponding corner diagram in Fig.~\ref{fig:corner_plot_radmcmc_all} illustrates the posterior distributions and parameter correlations of the four \texttt{RADEX} free parameters (kinetic temperature, atomic hydrogen density, electron density, and \ce{CH+} column density) obtained from the \texttt{RADEX/MCMC} sampling.
The first 2048 burn-in steps were discarded, and the posterior distributions were constructed from the subsequent 4096 production steps for each of the 32 walkers.

The posterior distributions are found to be well constrained and approximately Gaussian. In particular, the \ce{CH+} column density is found to be independent of the kinetic temperature and hydrogen atom density. In contrast, $N(\rm{CH^+})$ is anticorrelated with the electron density, reflecting the fact that it is primarily the product of $N(\rm{CH^+})$ 
and $n\rm{(e^-)}$ which is tightly constrained. In practice, the best model gives $T_{\rm kin} = 1258$~K, $n{\rm (H)}=1.62\times 10^5$~cm$^{-3}$, $n\rm{(e^-)}=5.89\times 10^3$~cm$^{-3}$, and $N[{\rm CH^+}]=1.48\times10^{14}$~cm$^{-2}$. The kinetic temperature and hydrogen density compare well with the average values derived from \textsc{Cloudy} around the \ce{CH+} plateau (see Table~\ref{tab:phy_ch+_peak}). In contrast, the electron density is higher by a factor of 10, while the \ce{CH+} column density is lower by a factor of $\sim 4$. As a result, the electron fraction, defined as $x_e=n\rm{(e^-)}/n({\rm H_{tot}})$ and approximated here by $x_e\sim n\rm{(e^-)}/n(\rm{H})$, is very high, $\sim 3.64\times10^{-2}$, and much higher than the expected value of $\sim 10^{-3}$ (i.e., the \ce{C+} abundance, since most electrons come from the ionization of atomic C; see Fig.~\ref{fig:abundance_density_profile}). On the other hand, the product of $N(\rm{CH^+})$ by $n\rm{(e^-)}$ agrees well with that calculated using the \textsc{Cloudy} values reported in Table~\ref{tab:phy_ch+_peak}, which supports the MCMC modeling. In addition, it should be noted that $n{\rm (e^-)}/n({\rm H})$ is an upper limit of the actual electron fraction, although the \ce{H_2} molecular fraction ($f_{\rm H_2}=2n({\rm H_2})/[n({\rm H})+2n(\rm{H_2})]$) at the \ce{CH+} abundance peak, as predicted by \textsc{Cloudy}, is low and typically between 1 and 10\% (see Table~\ref{tab:phy_ch+_peak}).

The predicted line fluxes are reported in \autoref{tab:line_fluxes_model_mcmc}.
The agreement between the MCMC best-fit and the observed \ce{CH+} line fluxes is better for ro-vibrational lines (to within a factor of $\exp{(\mu)}=0.97$ on average) than for rotational lines ($\exp{(\mu)}=0.50$) (see the red symbols in Fig~\ref{fig:CH+_line_ratio}). 
This result is reminiscent of what was reported previously. Our \textsc{Cloudy} results suggest that rotational and ro-vibrational lines arise from markedly distinct regions, differing in their temperature and spatial extension. \texttt{RADEX} calculations are therefore intrinsically limited to reproduce all lines (see Sect.~\ref{sec:radex}). The overpredicted electron density is likely a consequence of this limitation, namely, the use of a single-zone rather than a one-dimensional approach.
It should be noted that removing the ro-vibrational lines from the MCMC fit results in a much smaller electron fraction of $\sim 3\times 10^{-5}$, a much lower kinetic temperature of $\sim 200$~K, and a slightly higher \ce{CH+} column density of $\sim 4\times 10^{14}$~cm$^{-2}$. This also comes at the cost of a large underestimation of the ro-vibrational line fluxes. Alternatively, removing the rotational lines from the MCMC fit results in a lower \ce{CH+} column density of $\sim 3\times 10^{13}$~cm$^{-2}$, and does not substantially improve the electron fraction ($\sim 10^{-2}$). 

\begin{figure}
\centering
    \includegraphics[width=0.50\textwidth]{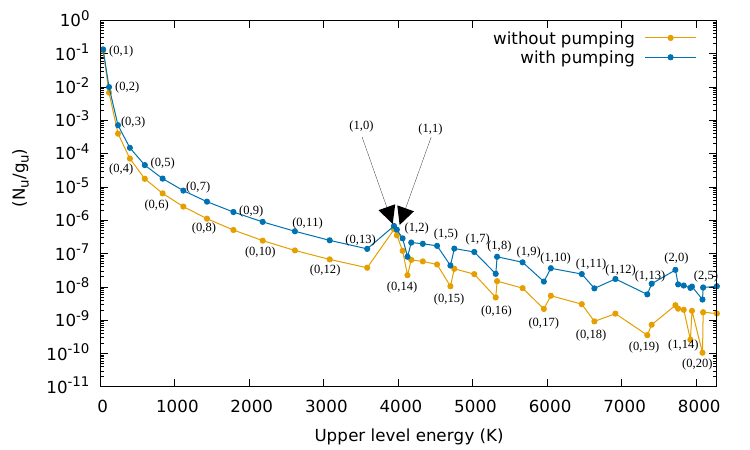}
    \caption{Ro-vibrational population diagram of \ce{CH+}~$(\upsilon,\ J)$ as predicted by \texttt{RADEX} considering the physical condition at ro-vibrational emission peak from the \textsc{Cloudy} model noted in \autoref{tab:phy_ch+_peak}.}
    \label{fig:ro-vib_pop}
\end{figure}

\begin{figure*}
     \includegraphics[width=0.49\textwidth]{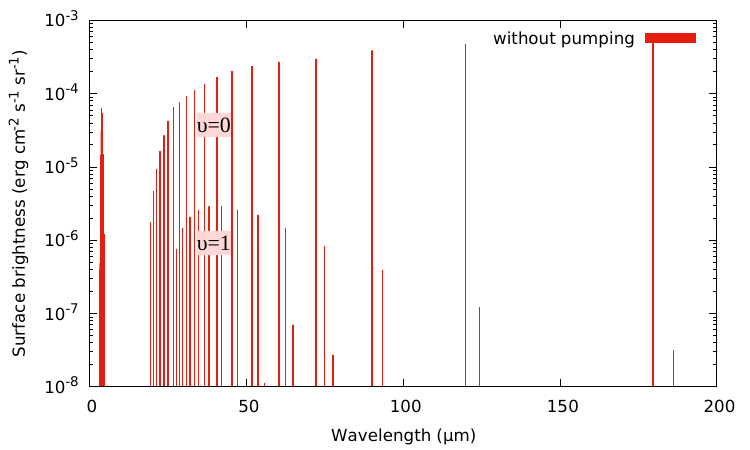}
     \includegraphics[width=0.49\textwidth]{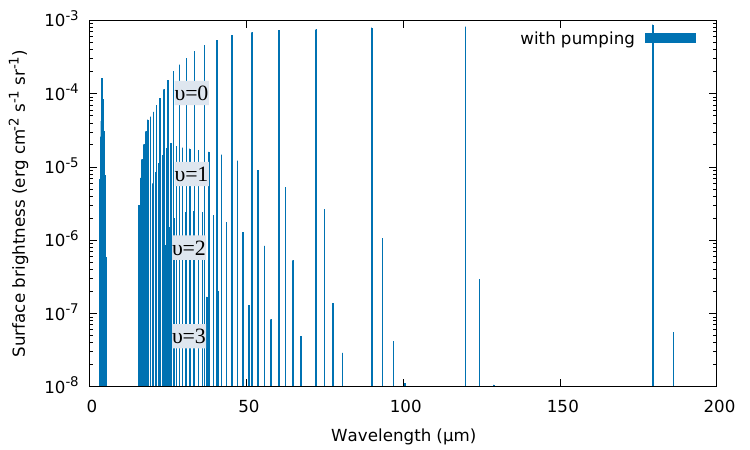}
     \includegraphics[width=0.49\textwidth]{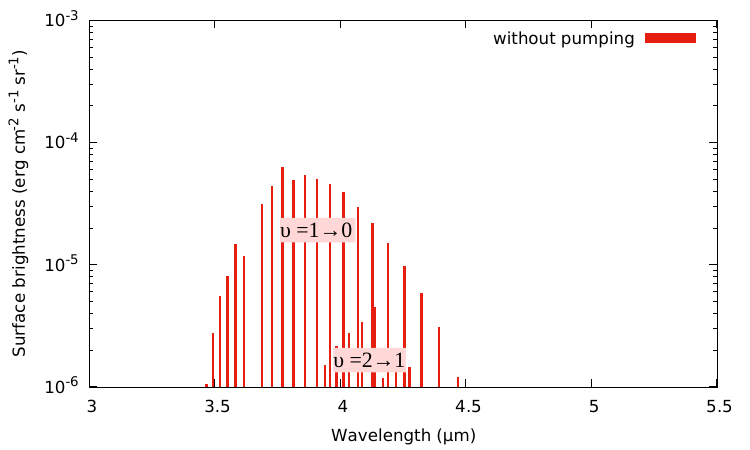}
     \includegraphics[width=0.49\textwidth]{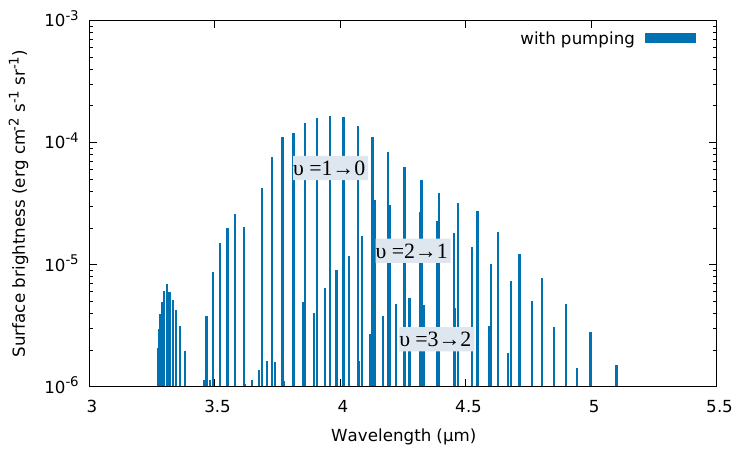}
     \caption{\ce{CH+} line surface brightness variation with wavelength as predicted by \texttt{RADEX} considering the physical condition at ro-vibrational emission peak from the \textsc{Cloudy} model noted in \autoref{tab:phy_ch+_peak}. The left panel shows this without chemical pumping, and the right panel considers chemical pumping.
     Upper panel: Full spectrum in the wavelength range $0 - 200$~\micron. Lower panel: Most intense ro-vibrational line spectrum in the wavelength range $3 - 5.5$~\micron. }
     \label{fig:ro-vib_lines}
 \end{figure*}

\subsection{Effect of chemical pumping}
\label{sec:excitation_comparison}

In order to assess the role of chemical pumping in the \ce{CH+} excitation, we compare below two \texttt{RADEX} models considering the physical conditions at the ro-vibrational \ce{CH+} emission peak from the \textsc{Cloudy} model (see \autoref{tab:phy_ch+_peak}): one model considers chemical pumping, and in the other, the chemical source and sink terms are ignored in the statistical-equilibrium equations. 
The results are illustrated in the population diagram of Fig.~\ref{fig:ro-vib_pop}, where the relative population (divided by the degeneracy) of levels with upper energy lower than 8274~K, that is, up to $(\upsilon=2, J=5$), is plotted as a function of the upper-level energy. Except for levels $(\upsilon=0, J=1)$ and to a lesser extent, $(\upsilon=1, J=0)$, the population of all levels is significantly affected by chemical pumping. This process is found to increase populations by up to a factor of 10. This can be rationalized by comparing the destruction and excitation rates of \ce{CH+}. For the approximate electron fraction $x_e \sim n({\rm e^-})/n({\rm H})=1.1 \times 10^{-3}$ (at the \ce{CH+} ro-vibrational emission peak; see Table~\ref{tab:phy_ch+_peak}), we expect hydrogen atoms and electrons to dominate the destruction and excitation of \ce{CH+}, respectively (see Appendices~\ref{sec:app_B} and \ref{sec:app_C}). As indicated by our collisional dataset, the destruction state-resolved rate coefficients for H are (at 1300~K) about $(4-9) \times 10^{-10}$~cm$^3$s$^{-1}$, while the electron-impact excitation rate coefficients from the ground $(\upsilon, J) = (0, 0)$ to the first excited level $(0, 1)$ is $1.2\times 10^{-6}$~cm$^3$s$^{-1}$, so that excitation of $(0, 1)$ dominates destruction. On the other hand, for higher levels, electron-impact excitation rate coefficients are lower than $\sim 4\times 10^{-7}$~cm$^3$s$^{-1}$ and destruction prevails. For these levels, the higher populations (with pumping) thus simply reflect their relatively short chemical lifetime. 
It should be noted that for levels higher than 8274~K, the chemical pumping effect boosts the level populations even further, by several orders of magnitude. 

More generally, chemical pumping of \ce{CH+} should be a major process in most PDR environments because the (state-resolved) destruction of \ce{CH+} by H or electrons is faster than or comparable to its excitation by these species, with rate coefficients at $T_{\rm kin}=1000$~K of $\sim 5\times 10^{-10}$~cm$^3$s$^{-1}$ and $\sim 3\times 10^{-7}$~cm$^3$s$^{-1}$ for destruction, compared to $\sim 10^{-10}$~cm$^3$s$^{-1}$ and $\sim 5\times 10^{-7}$~cm$^3$s$^{-1}$ for (de)excitation, respectively. In NGC~7027, electrons are the dominant exciting species of \ce{CH+} because the ionization fraction is large, but excitation by hydrogen atoms would prevail in other environments where $x_e\lesssim 10^{-4}$. Moreover, radiative excitation was found to play a minor role in NGC~7027 (see Sect.~\ref{sec:radex}). This can be understood by comparing the radiative pumping rate to the electron-impact excitation rate, for instance, for the transition $(0, 0) \rightarrow (1,1)$: with the radiative continuum flux from \textsc{Cloudy} at the corresponding wavelength of 3.615~\micron ($J_{\nu}\sim 1$~Jy/nsr), the infrared pumping rate is $\sim 2\times 10^{-9}$~s$^{-1}$, which is lower by two orders of magnitude than the electron-impact excitation rate of $4.1\times 10^{-7}$~s$^{-1}$ (at $T_{\rm kin}=1000$~K and $n(\rm e^-)=5.55\times 10^2$~cm$^{-3}$). Infrared pumping will thus significantly affect the \ce{CH+} population in environments with much stronger dust continua or much lower densities.

Finally, the line surface brightness variation with wavelength is plotted in Fig.~\ref{fig:ro-vib_lines}. The two spectra correspond to the populations presented in Fig.~\ref{fig:ro-vib_pop}, that is, with and without pumping, for physical conditions at the ro-vibrational emission peak from the \textsc{Cloudy} model.
\autoref{tab:CH+_transitions} presents the surface brightness of the most intense \ce{CH+} ro-vibrational transitions with and without pumping.
Chemical pumping increases the most intense lines (those with a surface brightness above $10^{-4}$~erg~cm$^{-2}$~s$^{-1}$~sr$^{-1}$) by factors of $\sim 2-3$. For weaker lines, much larger enhancement factors are observed, for instance, for pure rotational lines within $\upsilon=1$, which at $\sim 30$~\micron are predicted with line surface brightnesses up to $\sim 2\times 10^{-5}$~erg~cm$^{-2}$~s$^{-1}$~sr$^{-1}$ with pumping, but are lower by one order of magnitude without pumping. The detection of these transitions, as well as ro-vibrational lines of the $\upsilon=2 \to 1$ band at $\sim 4$~\micron, would provide an even more constraining diagnostic of the physical conditions. 
We note in this context that six ro-vibrational lines of the $\upsilon=2 \to 1$ band were recently detected with JWST toward the irradiated protoplanetary disk d203-506 (with line intensities of $(1-10)\times 10^{-6}$~erg~cm$^{-2}$~s$^{-1}$~sr$^{-1}$), where chemical pumping was investigated using a simple excitation model \citep{Zannese2025}.


\section{Discussion and conclusions} \label{sec:conclusions}

The exposure of dense, molecular gas to the UV radiation from the hot central star of the compact young PN NGC~7027 \citep[$T \simeq 198$~kK,][]{latt00} and to the X-ray emissions from shocked gas \citep{kastner01,montez18}, represents a rich laboratory for the interplay of radiative and chemical interactions. The conditions between the Str{\"o}mgren sphere and the inner boundary of the PDR provide a testbed for predictions of the production and destruction rates of reactive ions such as \ce{CO+} \citep{bubl23} and \ce{CH+} (this work). The latter contributes, through a series of two hydrogen abstraction reactions and thus bypassing the path through the slow radiative association of \ce{C+} and \ce{H2}, to the buildup of \ce{CH3+} \citep{smith92}, which in turn drives the organic chemistry in such environments. In this respect, the first detection of \ce{CH3+} in a planetary nebula \citep[NGC~6302,][with JWST/MIRI]{bhatt25} comes as no surprise, except for the fact that unlike NGC~7027, NGC~6302 is oxygen rich. Conversely, a product of oxygen chemistry, \ce{OH}, was recently and for the first time detected in a planetary nebula \citep{ouyang24}, which is none other than the carbon-rich NGC~7027. 

A deeper understanding of these observational findings requires simulations such as those presented in this study.
We used the 1D photoionization code \textsc{Cloudy} and the isothermal single-zone NLTE code \texttt{RADEX} (without and with chemical pumping) to conduct a comprehensive analysis aimed at determining how the observed \ce{CH+} emissions from NGC~7027 can help us to evaluate our understanding of the \ce{CH+} chemistry and excitation, and to place constraints on the physical conditions. The analysis was largely based on the recent availability of accurate sets of state-specific rate coefficients for the formation, excitation, and destruction of \ce{CH+}.

Although the limitations are similar as discussed in \citetalias{Sil2025}, our isobaric \textsc{Cloudy} model can reproduce observed \ce{CH+} pure rotational and ro-vibrational line intensities within a factor of 1.3 on average, with a maximum deviation of a factor of 3.
The predictions for \ce{HeH+} and H, \ce{He+} recombination lines are also entirely consistent with \citetalias{Sil2025}.
On the other hand, our stationary \textsc{Cloudy} model overpredicts the observed \ce{H2} emissions by up to a factor of 10, perhaps reflecting the short dynamical age of NGC~7027 and the time-dependent chemistry of \ce{H2}.
Our 1D \textsc{Cloudy} model also suggests that the rotational and ro-vibrational \ce{CH+} lines probe physically distinct regions of NGC~7027, with different kinetic temperatures.
Our \texttt{RADEX} model with chemical pumping, considering a uniform medium, is indeed unable to reproduce the rotational and ro-vibrational lines intensities simultaneously.
The overpredicted electron density (by a factor of $10$) obtained from the \texttt{RADEX/MCMC} best-fit model likely reflects the geometric limitation arising from the single-zone approximation.

Conversely, chemical pumping is found to markedly increase the populations of all \ce{CH+} levels above $(\upsilon=0$, $J=1)$.
Our \texttt{RADEX} model with chemical pumping thus suggests that the detection of pure rotational lines within $\upsilon=1$ and ro-vibrational lines of the $\upsilon=2\rightarrow1$ band would provide additional constraints on the physical conditions, especially the kinetic temperature gradient, just upstream of the \ce{H/H2} transition region.
We conclude that it is highly desirable to enable \textsc{Cloudy} to consistently couple subsets of the underlying chemical reaction network to the statistical equilibrium calculations.
It will also be interesting to implement the current set of state-resolved inelastic and reactive rates in a 1D complete PDR model, such as that of \cite{neuf21}.

\section*{Data availability}
The new extended \ce{CH+} ro-vibrational inelastic dataset underlying this article will be made available in the Excitation of Molecules and Atoms for Astrophysics (EMAA) database\footnote{\url{https://emaa.osug.fr} and \url{https://dx.doi.org/10.17178/EMAA}} \citep{Faure2025}.

\begin{acknowledgements}
The authors thank the anonymous referee for the valuable suggestions and constructive feedback, which have significantly improved the quality of this work.
M.S. acknowledges financial support from the European Research Council (consolidated grant COLLEXISM, grant agreement ID: 811363).
This research was carried out within the framework of IPAG, UMR 5274, with support from a CNRS postdoctoral fellowship.
M.S. further acknowledges support from the National Science and Technology Council, Taiwan (Grant Nos. NSTC-114-2811-M-007-017, NSTC-111-2112-M-007-014-MY3, NSTC-113-2639-M-A49-002-ASP, and NSTC-113-2112-M-007-027), and from the Postdoctoral Incentive Award of the Research Talent Resource Center at National Tsing Hua University.
T.G.L. acknowledges funding from Grant No. MICIU/AIE/10.13039/501100011033 
PID2024-155666NB-I00.
J.F acknowledges support from the USA National Science Foundation award number AST-2303895.
This research made use of \textsc{Cloudy} \citep{chat23}, \texttt{RADEX} \citep{vand07}, \texttt{emcee} \citep{foreman2013}, \texttt{astropy} \citep[\url{http://www.astropy.org};][]{Astropy_2022}, \texttt{corner} \citep{foreman2016}, \texttt{matplotlib} \citep{Hunter2007}, \texttt{numpy} \citep{Harris2020}, \texttt{UKRmol+ 3.2} \citep[for the electron-\ce{CH+} scattering calculations;][]{Masin2020}, and \texttt{Psi4} \citep[for the underlying \ce{CH+} electronic structure calculations;][]{Smith2020}.
This document was prepared using the collaborative tool Overleaf available at: \url{https://www.overleaf.com/}.

\end{acknowledgements}

\bibliographystyle{aa}
\bibliography{references}

\begin{appendix}

\section{Updated reaction rates} \label{sec:app_A}

The \ce{CH+} chemical network from \textsc{Cloudy} version c23.01 was updated for five among the six dominant formation and destruction reactions R1$-$R6, as listed in Table~\ref{tab:updated_reaction}. For the hydrogenation of C$^+(^2P)$ by \ce{H2} (reaction R1), we follow \textsc{Cloudy}'s distinction between two forms of molecular hydrogen: \ce{H2} represents the ground vibrational state ($\upsilon=0)$ and H$_2^*$ refers to vibrationally excited states ($\upsilon \geq 1$).
\textsc{Cloudy} defines H$_2^*$ as all levels in the $X$ electronic state with an energy greater than 4100 cm$^{-1}$.
In practice, we assumed that H$_2$ corresponds to the first 7 ro-vibrational pure rotational levels ($\upsilon=0, J=0\to6$), whose energies are significantly below the endothermicity $\Delta H_0^{\rm o}=4620$~K of the reaction C$^+(^2P)$ + H$_2(^1\Sigma^+) \to$ CH$^+(X^1\Sigma^+)$ + H$(^2S)$ \citep{Hierl1997}. Higher levels were attributed to H$_2^*$, that is, the form of molecular hydrogen which is reactive with C$^+(^2P)$. A uniform excitation temperature of 1900~K was adopted for all levels of molecular hydrogen, which was found to reproduce the rotational diagram for the observed transitions of H$_2$ fairly well \citep{neuf21}. The state-to-state rate coefficients of \cite{Gonzalez2026} (see Appendix~\ref{sec:app_B}) were finally employed to derive the rate coefficients for H$_2$ and H$_2^*$ listed in Table~\ref{tab:updated_reaction}. For the photoionization of CH (reaction R2), we employed the $A_{\rm V}$-dependent rate coefficient from \cite{Heays2017} with two different product channels as recommended by KIDA\footnote{\url{https://kida.astrochem-tools.org/}}. For the dissociative recombination (DR) of \ce{CH+} (reaction R3), the rate coefficient measured by \cite{Paul2022} for the ground state of CH$^+(\upsilon=0, J=0)$ was used. It should be very similar to the thermal rate coefficient given the weak $J$-specific dependence of the DR process (see Appendix~\ref{sec:app_B}). For the reaction of \ce{CH+} with H (reaction R4), we employed the thermal time-independent quantum mechanical (TIQM) rate coefficient computed by \cite{Faure2017}, which agrees well with available experimental measurements above 50~K (see Appendix~\ref{sec:app_B}). For the reaction of \ce{CH+} with H$_2$ (reaction R5), the default \textsc{Cloudy} value was employed, as recommended by KIDA and UMIST\footnote{\url{https://umistdatabase.uk/}}. Finally, for the photodissociation of \ce{CH+} (reaction R6), the $A_{\rm V}$-dependent rate coefficient was taken from \cite{Heays2017}. For the reactions R1, R3, and R4, the rate coefficients were fitted using a standard modified-Arrhenius equation. We note that our fits are reliable over the kinetic temperature ranges $100-3000$~K for reaction R1, $10-10\,000$~K for reaction R3, and $10-3000$~K for reaction R4, and they should be accurate to within $\sim$ 20\%.
We further point out that \citetalias{Sil2025} contains a typographical error in the thermal rate coefficient for reaction R1 between \ce{He+} and H; the corrected form is noted here in \autoref{tab:updated_reaction}.

\begin{table*}[h!]
\centering
\caption{Rate coefficient of dominant formation and destruction reactions of \ce{CH+} adopted in the \textsc{Cloudy} chemical model.}\label{tab:updated_reaction}
\resizebox{\linewidth}{!}{\begin{tabular}{|c|c|c|c|c|}
 \hline
 \hline
{\bf Reaction No. (Type)} & {\bf Reactions} & {\bf Rate coefficients} & {\bf Units} & {\bf References} \\
\hline
\hline
\multicolumn{5}{|c|}{\bf Formation reactions} \\
\hline
R1 (IN) & $\rm{C^+ + H_2^* \rightarrow CH^+ + H}$ & $6.0\times10^{-11}$ & cm$^3$s$^{-1}$ &  This work \\
 & $\rm{C^+ + H_2 \rightarrow CH^+ + H}$ & $\rm{6.7\times10^{-11}~exp(-1971/T)}$ & cm$^3$s$^{-1}$ & This work \\
\hline
R2 (PH) & CH + $h\nu \rightarrow$ \ce{CH+} + e$^-$ & $\rm{7.6\times10^{-10}~exp(-3.67~A_V)}$ & s$^{-1}$ & \cite{Heays2017} (KIDA BR = 46\%) \\
& CH + $h\nu \rightarrow$ C + H & $\rm{9.1\times10^{-10}~exp(-2.12~A_V)}$ & s$^{-1}$ & \cite{Heays2017} (KIDA BR = 54\%) \\
\hline
\multicolumn{5}{|c|}{\bf Destruction reactions} \\
\hline
R3 (DR) & $\rm{CH^+ + e^- \rightarrow C + H}$ & $\rm{1.7\times10^{-7}~(T/300)^{-0.75}}$ & cm$^3$s$^{-1}$ & This work  \\
\hline
R4 (IN) & $\rm{CH^+ + H \rightarrow C^+ + H_2}$ & $\rm{1.0\times10^{-9}~(T/300)^{-0.45}~exp(-12.4/T)}$ & cm$^3$s$^{-1}$ & This work \\
\hline
R5 (IN) & $\rm{CH^+ + H_2 \rightarrow CH_2^+ + H}$ & $\rm{1.20\times10^{-9}}$ & cm$^3$s$^{-1}$ & \textsc{Cloudy} default \\
\hline
R6 (PH) & \ce{CH+} + $h\nu \rightarrow$ C + \ce{H+} 
 & $\rm{3.3\times10^{-10}~exp(-3.54~A_V)}$ & s$^{-1}$ & \cite{Heays2017} \\
\hline
\hline
\multicolumn{5}{|c|}{\bf \ce{HeH+} formation reaction} \\
\hline
R1* (RA) & \ce{He+} + H $\rightarrow$ \ce{HeH+} + $h\nu$ & $\rm{2.32\times10^{-16}~exp(122/T)}$ & cm$^3$s$^{-1}$ & \citetalias{Sil2025} \\
\hline
\hline
\end{tabular}}
\tablefoot{
IN refers to ion-neutral reactions, PH to photodissociation reactions, DR to dissociative recombination reactions, RA to radiative association reactions, and h$\nu$ to a photon energy. \\
(*) The thermal rate coefficient for reaction R1 (\ce{He+} + H $\rightarrow$ \ce{HeH+} + $h\nu$) corrects a typographical error in \citetalias{Sil2025}.}
\end{table*}

\section{New \ce{CH+} ro-vibrational data for reactive collisions} \label{sec:app_B}

This dataset was employed in the \texttt{RADEX/MCMC} calculations (see Sect.~\ref{sec:mcmc}).

\subsection{Hydrogenation of C$^+(^2P)$} \label{sec:RA}

\begin{figure}[h]
    \includegraphics[width=0.45\textwidth]{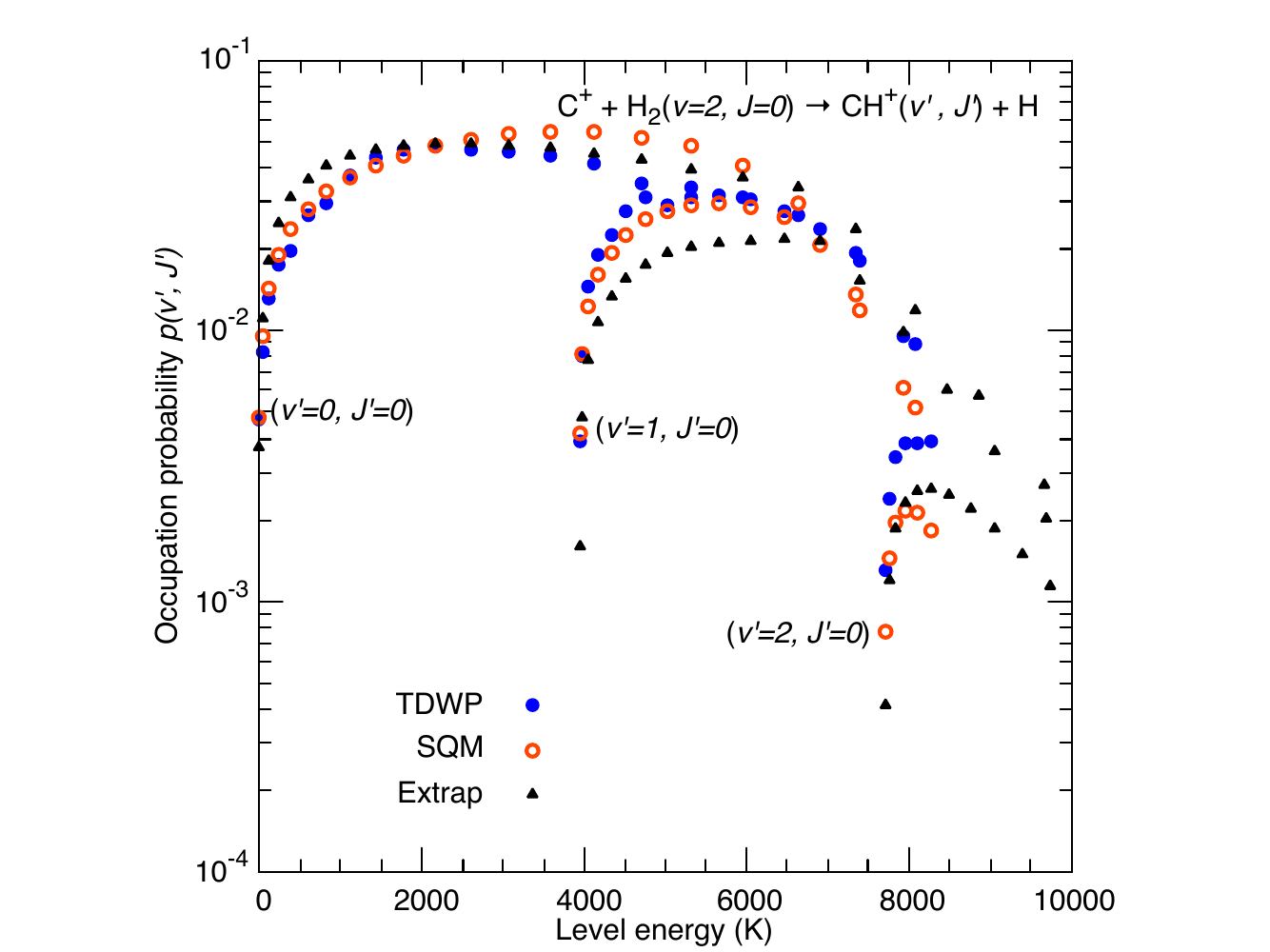}
    \caption{Occupation probability of the nascent \ce{CH+} product from the reaction of C$^+(^2P)$ with H$_2$ in the ro-vibrationally excited state $(\upsilon, J)=(2, 0)$ as function of the CH$^+(\upsilon', J')$ level energy and at a kinetic temperature $T_{\rm kin} = 1000$~K. The SQM calculations of \cite{Gonzalez2026} are compared to the TDWP results of \cite{Faure2017} and to the simple extrapolation scheme (Extrap) defined by Eq.~\ref{eqn:B1}. See text for details. The rotational ground states in the rotational manifold of $\upsilon'=0, 1, 2$ are labeled for clarity.}
    \label{fig:popchp}
\end{figure}

For the reaction of C$^+(^2P)$ with H$_2(\upsilon, J)$, which produces CH$^+(\upsilon', J')$ + H, we employed the extensive set of rate coefficients computed recently by \cite{Gonzalez2026} using the statistical quantum method (SQM). The use of statistical methods is justified for this system because the intermediate complex \ce{CH2+} forms a deep potential well \citep{Konings2021,Gonzalez2026}. The SQM calculations were performed for the first 31 levels of H$_2$ (up to $(\upsilon, J)=(1, 10)$ at 14\,223~K above the ground state) and for the first 42 levels of \ce{CH+} (up to $(\upsilon', J')=(2, 5)$ at 8275~K above the ground state). They were compared to results obtained with a different statistical approach \citep{Konings2021} and to the results of \cite{Faure2017}, based on the more accurate but expensive time-dependent wave-packet (TDWP) method, and a good agreement (within a factor of 2) was observed at the state-to-state level. This is illustrated in Fig.~\ref{fig:popchp} where we have plotted the occupation probability of \ce{CH+} following the reaction of C$^+(^2P)$ with the excited state of H$_2$ $(\upsilon', J')=(2, 0)$ at $T_{\rm kin}=1000$~K\footnote{The occupation probability, $p(\upsilon', J')$, is directly proportional to the state-to-state rate coefficient (in cm$^3$s$^{-1}$) for the reaction C$^+(^2P)$ + H$_2(\upsilon=2, J=0)\to$ CH$^+(\upsilon', J')$ + H.}. We note that a good agreement was also observed between the SQM calculations and experiments performed with H$_2(\upsilon=0, 1)$ at different temperatures, see e.g., Fig.~1 of \cite{Gonzalez2026}. The full set of SQM rate coefficients includes 1194 state-to-state transitions and covers the kinetic temperature range from 30 to 3000~K. We note that the original reactive rate coefficients of \cite{Gonzalez2026} are limited to the temperature range $30-1500$~K to ensure full convergence. They were extended here up to 3000~K to better cover the physical conditions of NCG~7027, with limited impact on accuracy \citep[see][]{Gonzalez2026}.

In order to cover all 80 \ce{CH+} levels in its ground electronic state $X^1\Sigma^+$, we have also extrapolated the data of \cite{Gonzalez2026} (that is, above $(\upsilon', J')=(2, 5)$) using a simple formulation similar to that of \cite{neuf21} (see their Appendix~B.2), where a formation temperature is introduced additionally. The (normalized) probability $p(\upsilon', J')$ of forming \ce{CH+} in a level $(\upsilon', J')$ is defined as:
\begin{equation}
p(v', J')=
\begin{cases}
\frac{(2J+1)\exp(-E/T_{\rm form})}{\sum (2J+1)\exp(-E/T_{\rm form})}  & \text{if } E < 0 \\[2ex]
\frac{(2J+1)\exp(-E/T_{\rm kin})}{\sum (2J+1)\exp(-E/T_{\rm kin})}  & \text{if } E \geq 0,
\end{cases}
\label{eqn:B1}
\end{equation}
where $E=\Delta H_0^{\rm o} + E_{{\rm CH^+}}(\upsilon', J') - E_{{\rm H_2}}(\upsilon, J)$ is the activation energy, including the reaction endothermicity $\Delta H_0^{\rm o}=4620$~K \citep{Hierl1997} and $E_{{\rm CH^+}}(\upsilon', J')$ and $E_{{\rm H_2}}(\upsilon, J)$ are the ro-vibrational energies of \ce{CH+} and H$_2$, respectively. The formation temperature $T_{\rm form}$ was taken as $2/3\times(E_{{\rm H_2}}(v, J)-\Delta H_0^{\rm o})$, as observed and recommended by \cite{Faure2017}. We note that in the measurements of the deuterated variants of the H$_2$ + H$_2^+$ reaction, about two-thirds of the reaction exothermicity was also found in the internal ro-vibrational energies of the (triatomic) product ions \citep[][although their population does not necessarily follow a Boltzmann distribution]{Merkt2022}. As can be seen in Fig.~\ref{fig:popchp}, the above extrapolation scheme {(Extrap)} reproduces the SQM probabilities to within a factor of 2 for the first 42 levels of \ce{CH+}.

\begin{figure}[h]
    \includegraphics[width=0.45\textwidth]{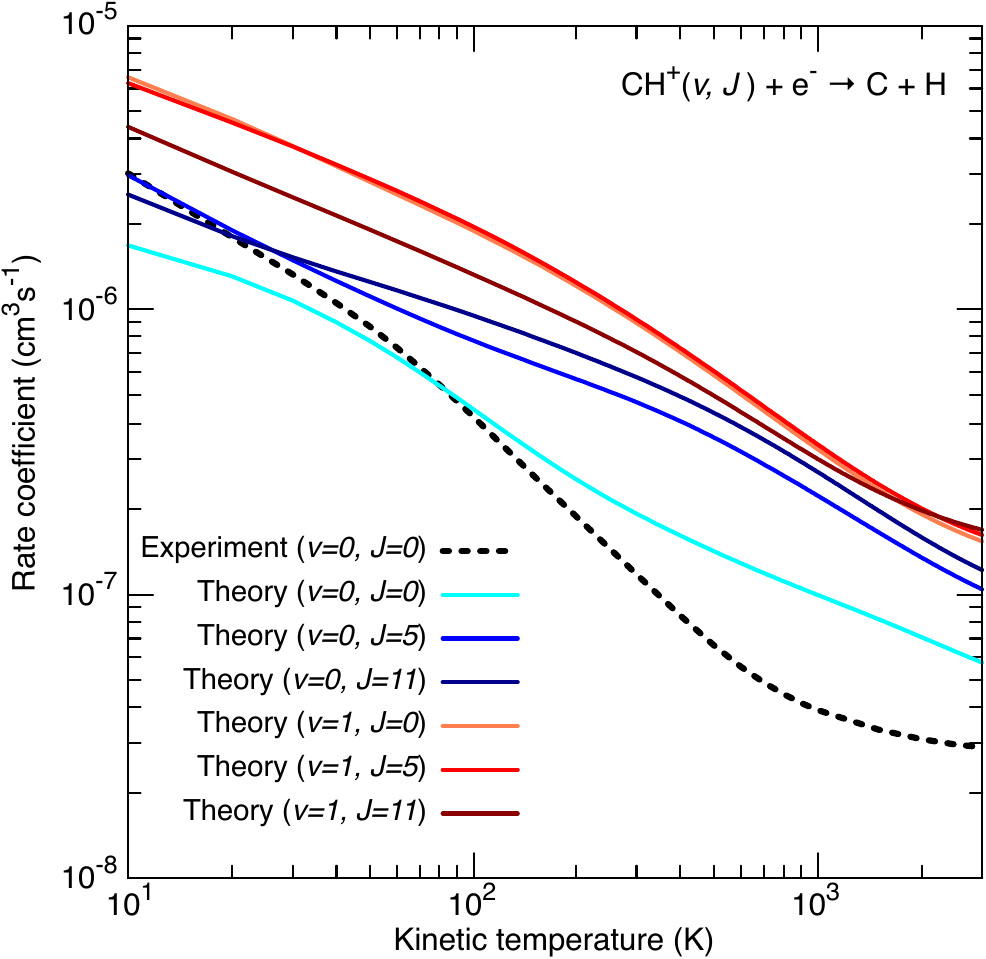}
    \caption{State-selected DR rate coefficients of \ce{CH+}($\upsilon, J$). The present MQDT calculations are denoted by colored solid lines. The thick black dashed line gives the experimental rate coefficients from \cite{Paul2022} for $J=0$.}
    \label{fig:dr-rate}
\end{figure}

\subsection{Dissociative recombination of \ce{CH+} with electrons} \label{sec:DR}

For the DR of $\ce{CH+}(v,J)$ with electrons, we have computed new cross sections using the theoretical treatment of \citet{forer2024} combining the R-matrix method with the multichannel quantum defect theory (MQDT), adopting here a partial wave basis $l=0-4$ and electron energies up to 1~eV. Rate coefficients were obtained for the \ce{CH+} levels $(\upsilon=0, J=0-11)$ and $(\upsilon=1, J=0-11)$. As in \cite{forer2024}, these rate coefficients could be compared to state-resolved measurements made at the Cryogenic Storage Ring (CSR) by \cite{Paul2022}. These authors were able to determine the $J$-specific rate coefficient for the levels $J=0,1$ of \ce{CH+} in its ground electronic and vibrational state, and for kinetic temperatures between 10 and 40\,000~K. Our new DR rate coefficients for $\upsilon=0$, $J=0-1$ agree with those of \cite{Paul2022} to within a factor of $\sim2-3$ in the range $10-3000$~K, as illustrated in Fig.~\ref{fig:dr-rate} for $J=0$. We can also observe that the rotational effects are weak, with $J$-specific DR rate coefficients for \ce{CH+}($\upsilon=0$) differing by less than a factor of 3 at 1000~K. Moreover, the DR rate coefficients of \ce{CH+}($\upsilon=1$) are found to exceed those of \ce{CH+}($\upsilon=0$) by a small factor ($<3$), and rotational effects are even less significant. In practice, the MQDT rate coefficients for \ce{CH+}($\upsilon=0, J=11$) were used for all levels ($\upsilon=0, J>11$) and the rate coefficients for ($\upsilon=1, J=11$) were employed for all levels ($\upsilon=1, J>11$) and ($\upsilon>1, J$). We thus selected only theoretical DR rate coefficients, rather than the experimental ones for ($\upsilon=0, J=0-1$), to ensure consistency among the ($\upsilon, J$)-specific DR rate coefficients and with the excitation rate coefficients described below.

\subsection{The reaction of \ce{CH+} with H atoms}

For the reaction of \ce{CH+} with H (which produces C$^+(^2P)$ + H$_2$), we used TIQM calculations of \cite{Faure2017} for \ce{CH+} in levels $(\upsilon=0, J=13)$ and the SQM calculations of \cite{Gonzalez2026} for higher levels up to ($\upsilon=2, J=5$), consistent with the calculations described above for the reverse reaction between C$^+(^2P)$ and H$_2$. As for this latter reaction, this set of state-resolved rate coefficients (summed over the final ro-vibrational states of H$_2$) was found to be in good agreement with the more accurate TIQM calculations of \cite{Faure2017}, typically within a factor of 2. The SQM thermal rate coefficient thus matches well with the thermal TIQM value and also with various measurements, except below $\sim 50$~K where a strong decrease of the rate coefficient was observed in the ion-trap experiment of \cite{Plasil2011}. This surprising result was interpreted by these authors as a dramatic loss of reactivity of rotationally cold \ce{CH+} ions at very low temperatures. As recently discussed by \cite{delMazo2025}, the vibrational excitation of \ce{CH+} strongly reduces reactivity at low temperatures, so that vibrational thermalization was perhaps not fully achieved in the ion-trap experiment \citep[see Fig.~9 of][]{delMazo2025}. A similar result was observed by \cite{Gonzalez2026}. For levels not included in the SQM set of \cite{Gonzalez2026}, that is, for levels above CH$^+(\upsilon=2, J=5)$, we used an average temperature-independent rate coefficient of $4\times 10^{-10}$~cm$^3$s$^{-1}$. We note that rotational and vibrational effects are weak for kinetic temperatures above $\sim 500$~K, with ro-vibrational rate coefficients lying in the range $(4-5)\times 10^{-10}$~cm$^3$s$^{-1}$. This range is $\sim 2-3$ orders of magnitude lower than the above DR rate coefficients so that hydrogen atoms will dominate the destruction of \ce{CH+} in regions where $n({\rm e})/n({\rm H})< 2\times 10^{-3}$.

\section{New \ce{CH+} ro-vibrational data for inelastic collisions} \label{sec:app_C}

This dataset was employed in the \textsc{Cloudy} and \texttt{RADEX} calculations (see Sect.~\ref{sec:NLTE_cloudy} and \ref{sec:radex}).

\subsection{Electron collisions}

For inelastic collisions of CH$^+(\upsilon, J)$ with electrons, new cross sections were computed using the theoretical treatment of \cite{forer2024}, adopting a partial wave basis $l=0-4$ and electron energies up to 1~eV, as for DR. Pure rotational and ro-vibrational rate coefficients were obtained for the \ce{CH+} levels $(\upsilon=0, J=0-11)$ and $(\upsilon=1, J=0-11)$ --- all calculated as described by \citet{forer2024}. This method includes the lowest two electronically excited states of the ion, which produce low-energy resonances in the scattering phases of the ground electronic state and can significantly increase the vibrational coupling between different vibrational levels. For transitions with $\Delta J=\pm 1$, the limited partial-wave basis of the scattering calculations is corrected via Born closure with the Coulomb-Born (CB) approximation, as in \cite{hami16}. The CB approximation accounts for the larger partial waves that are omitted in the R-matrix scattering basis; these are important due to strong dipole-induced coupling between partial waves that satisfy $\Delta l=\pm1$.
The CB approximation is known to be reliable for strongly dipolar ions like \ce{CH+} (1.68~D), but does not include resonances in the scattering phases and, therefore, is only used to supplement the limited partial-wave basis of the R-matrix calculations. In fact, the CB approximation on its own was found to underestimate the ro-vibrational rate coefficients by three orders of magnitude overall. 

\begin{figure}
    \includegraphics[width=0.45\textwidth]{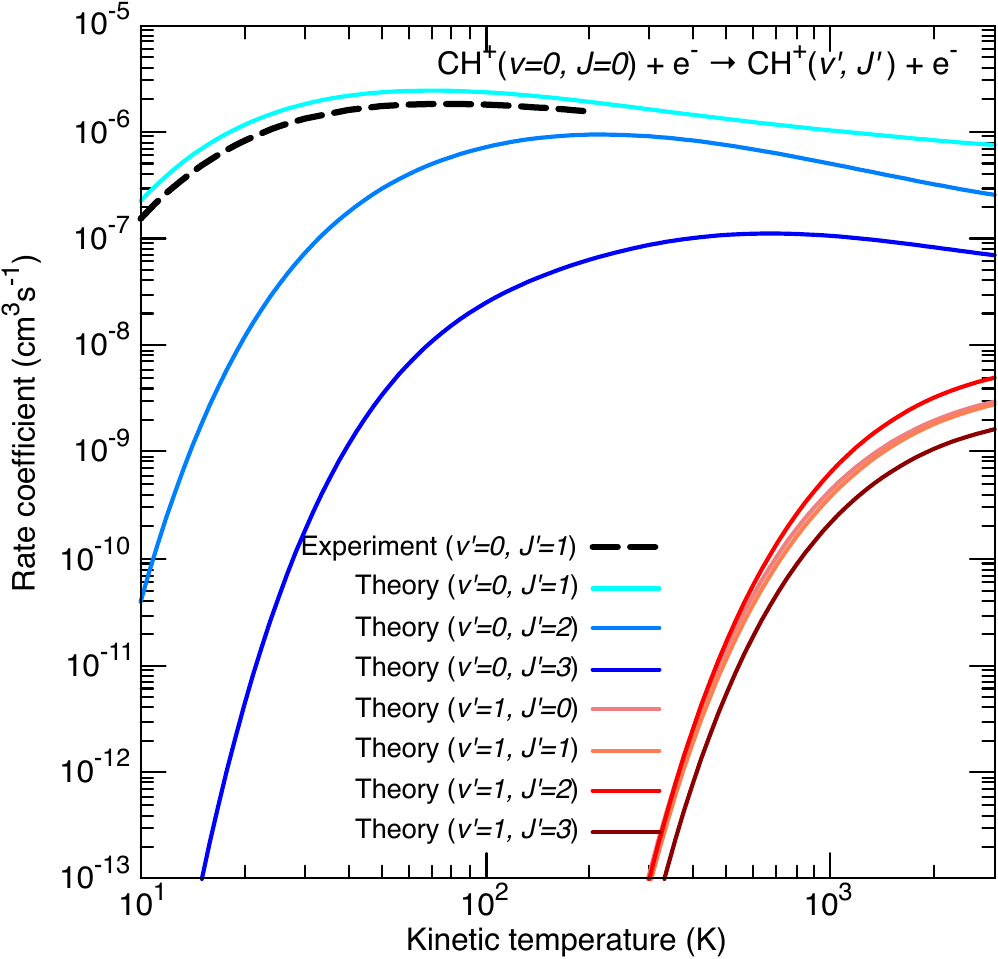}
    \caption{State-to-state excitation rate coefficients from the ground state of \ce{CH+}($\upsilon=0, J=0$) to the lowest three and four rotational and ro-vibrational excited states ($\upsilon', J'$), respectively. The present MQDT calculations (CB corrected) are denoted by colored solid lines. The thick black dashed line gives the experimental rate coefficients from \cite{Kalosi2022} for $\upsilon'=0, J'=1$.}
    \label{fig:ex-rate}
\end{figure}

As in \cite{forer2024}, our pure rotational rate coefficients could be compared to state-selected measurements made at the CSR by \cite{Kalosi2022}. These authors were able to evaluate the rate coefficients for the excitation $J=0 \to J=1$ of \ce{CH+} in its ground electronic and vibrational state, for kinetic temperatures up to 200~K. Our new excitation rate coefficients agree with those of \cite{Kalosi2022} to within 50\% across the $10-200$~K temperature, as illustrated in Fig.~\ref{fig:ex-rate}, which is consistent with the experimental one-sigma uncertainty. We can also observe that the dominant transitions are those with $\Delta J=1, 2$ for rotational excitation and $\Delta J=0, 1, 2$ for ro-vibrational excitation. Another important result is that ro-vibrational excitation is dominated by transitions with $\Delta \upsilon=1$ \citep[see Fig.~2 of][]{forer2024} and that the corresponding rate coefficients do not significantly depend on the initial state $\upsilon$, especially above 100~K. As a result, the pure rotational rate coefficients for de-excitations $\Delta J=-1, -2$ of \ce{CH+}($\upsilon=0, J=11$) and ($\upsilon=1, J=11$) were used for all levels ($\upsilon=0, J>11$), ($\upsilon=1, J>11$) and  ($\upsilon\geq 1, J$), respectively. Similarly, the ro-vibrational rate coefficients for (de-)excitations $\Delta \upsilon=-1; \Delta J=0, \pm 1, \pm 2$ of \ce{CH+}($\upsilon=1, J=9$) were used for all levels ($\upsilon=1, J>11$) and ($\upsilon>1, J$). The full set of \ce{CH+ + e-} inelastic rate coefficients is available for temperatures between 10 and 3000~K.

\subsection{Collision with hydrogen atoms}

For inelastic collisions of CH$^+(\upsilon, J)$ with hydrogen atoms, we used the TIQM calculations of \cite{Faure2017} for CH$^+(\upsilon=0, J=0-13)$ and the SQM rate coefficients computed by \cite{Gonzalez2026} for levels above $(\upsilon=2, J=5)$, consistent with the calculations used for the \ce{CH+} + H reactive channel described above. We note that since \ce{CH+} + H has a statistical behavior \citep[][and references therein]{Gonzalez2026}, inelastic collisional propensity rules are different from non-statistical systems like, e.g., HeH$^+$ \citepalias[][and references therein]{Sil2025}. In particular, for a specific initial ro-vibrational state ($\upsilon, J)$, the ro-vibrational rate coefficients for a transition $(\upsilon, J)\to (\upsilon', J')$ do not depend on $\upsilon'$ but only on $J'$. Thus, for levels higher than $(\upsilon=2, J=5)$, we extrapolated the data of \cite{Gonzalez2026} by applying a temperature-independent de-excitation rate coefficients of $5\times 10^{-11}$cm$^3$s$^{-1}$ for all rotational and ro-vibrational transitions with $|\Delta J|\le 5$. This crude prescription should be accurate to within a factor of $5-10$. We note that the rate coefficients for the rotational (ro-vibrational) excitation of \ce{CH+} due to H collisions are typically four (three) orders of magnitude smaller than those due to electron collisions, meaning that hydrogen atoms will dominate the \ce{CH+} excitation if $n({\rm e})/n({\rm H}) < 10^{-4}$. The full set of \ce{CH+} + H inelastic rate coefficients is available for temperatures between 30 and 3000~K. We note that the original inelastic rate coefficients of \cite{Gonzalez2026} are limited to the temperature range $30-1500$~K to ensure full convergence. They were extended here up to 3000~K to better cover the physical conditions of NCG~7027, with limited impact on accuracy \citep[see][]{Gonzalez2026}.    

\section{Emissivity profiles of \ce{CH+} and \ce{H2}}
\begin{figure}
    \centering
    \includegraphics[width=\linewidth]{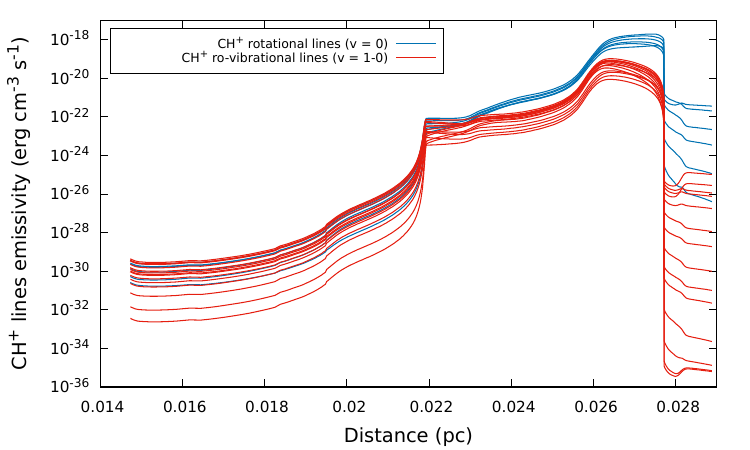}
    \caption{Emissivity profile of \ce{CH+} pure rotational lines and ro-vibrational lines as a function of distance.}
    \label{fig:emissivity_CH+}
\end{figure}

\begin{figure}
    \centering
    \includegraphics[width=\linewidth]{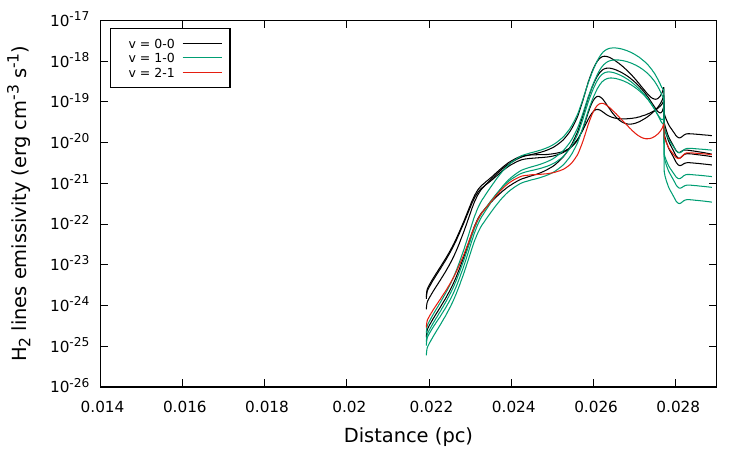}
    \caption{Emissivity profile of the observed \ce{H2} lines as a function of distance.}
    \label{fig:emissivity_H2}
\end{figure}

Figures~\ref{fig:emissivity_CH+} and \ref{fig:emissivity_H2} display the intrinsic line emissivities of \ce{CH+} and \ce{H2}, respectively, as a function of the distance from the star.

\section{\texttt{RADEX/MCMC} fitting results} \label{sec:app_D}

\begin{figure}
    \includegraphics[width=0.5\textwidth]{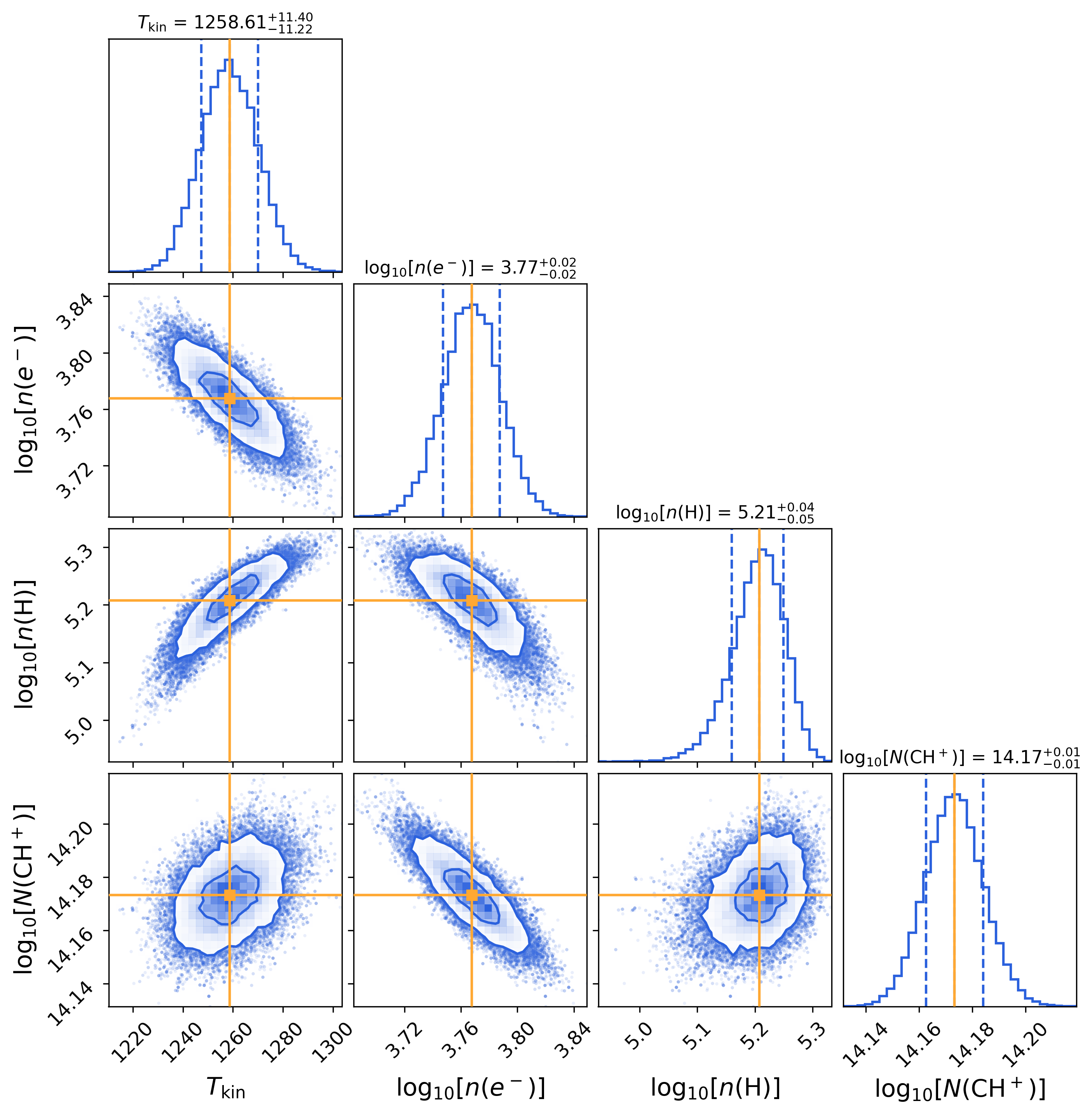}
    \caption{Corner plot obtained from the \texttt{RADEX/MCMC} exploration for \ce{CH+}, taking into account all detected rotational and ro-vibrational transitions. The diagonal panels show the marginalized posterior probability distributions of $T_{\rm kin}$ (linear scale),  $n(\rm e^-)$, $n({\rm H})$, and $N(\rm{CH^+})$ ($\log_{10}$ scale), while the off-diagonal panels display the corresponding pairwise correlations. The orange lines and markers indicate the maximum posterior estimate, and the vertical dashed lines in each diagonal panel mark the 16th and 84th percentiles (i.e., the $1\sigma$ credible intervals). The contours in the 2D panels enclose 39.3\% and 86.4\% of the posterior probability, corresponding to the $1\sigma$ and $2\sigma$ confidence regions for a two-dimensional Gaussian distribution.}
    \label{fig:corner_plot_radmcmc_all}
\end{figure}

The corner diagram resulting from the \texttt{RADEX/MCMC} analysis of \ce{CH+} line fluxes is presented in Fig.~\ref{fig:corner_plot_radmcmc_all}.

\section{\ce{CH+} line surface brightness prediction}

 The surface brightness values of the strongest \ce{CH+} ro-vibrational transitions in the presence and absence of chemical pumping are noted in \autoref{tab:CH+_transitions}.
 The values are predicted by \texttt{RADEX} considering the physical condition at the ro-vibrational emission peak and \ce{CH+} total column density obtained from the \textsc{Cloudy} model noted in \autoref{tab:phy_ch+_peak}.
 A line width (FWHM) of 30~km~s$^{-1}$ and the incident radiative continuum at the illuminated face of the cloud, obtained from the \textsc{Cloudy} model, were used as the background radiation field.

\begin{table*} 
\caption{Line surface brightness of the most intense ro-vibrational transitions of \ce{CH+} with and without considering chemical pumping as predicted by \texttt{RADEX} considering the physical condition at ro-vibrational emission peak from the \textsc{Cloudy} model noted in \autoref{tab:phy_ch+_peak}.
\label{tab:CH+_transitions}}
\centering
\begin{tabular}{ccccc}
\hline
\hline
{\bf \ce{CH+} lines} & {\bf Wavelength} & {\bf Upper state} & \multicolumn{2}{c}{\bf Surface Brightness (erg~cm$\bm{^{-2}}$~s$\bm{^{-1}}$~sr$\bm{^{-1}}$)} \\
$\bm{(\upsilon^\prime, J^\prime)\to(\upsilon, J)}$ & {\bf (\micron)} & {\bf energy / $\bm{k_B}$ (K)} & {\bf Without pumping} & {\bf With pumping} \\
\hline
\hline
\multicolumn{5}{c}{$\bm{\upsilon=1\rightarrow0}$} \\
\hline
$(1, 0) \rightarrow (0,1)\ [P(1)]$ & 3.6876 & 3942 & $3.14\times10^{-5}$ & $4.22\times10^{-5}$  \\
$(1, 1) \rightarrow (0,0)\ [R(0)]$ & 3.6146 & 3980 & $1.18\times10^{-5}$ & $2.04\times10^{-5}$ \\
$(1, 1) \rightarrow (0,2)\ [P(2)]$ & 3.7272 & 3980 & $4.36\times10^{-5}$ & $7.52\times10^{-5}$ \\
$(1, 2) \rightarrow (0,3)\ [P(3)]$ & 3.7689 & 4058 & $6.27\times10^{-5}$ & $1.10\times10^{-4}$ \\
$(1, 3) \rightarrow (0,4)\ [P(4)]$ & 3.8129 & 4174 & $4.89\times10^{-5}$ & $1.20\times10^{-4}$ \\
$(1, 4) \rightarrow (0,5)\ [P(5)]$ & 3.8591 & 4328 & $5.38\times10^{-5}$ & $1.45\times10^{-4}$ \\
$(1, 5) \rightarrow (0,6)\ [P(6)]$ & 3.9078  & 4520 & $5.01\times10^{-5}$ & $1.58\times10^{-4}$ \\
$(1, 6) \rightarrow (0,7)\ [P(7)]$ & 3.9589 & 4751 & $4.58\times10^{-5}$ & $1.64\times10^{-4}$ \\
$(1, 7) \rightarrow (0,8)\ [P(8)]$ & 4.0125 & 5019 & $3.93\times10^{-5}$  & $1.60\times10^{-4}$ \\ 
$(1, 8) \rightarrow (0,9)\ [P(9)]$ & 4.0688 & 5324 & $2.94\times10^{-5}$ & $1.37\times10^{-4}$ \\
$(1, 9) \rightarrow (0,10)\ [P(10)]$ & 4.1278 & 5667 & $2.17\times10^{-5}$ & $1.10\times10^{-4}$ \\
$(1, 10) \rightarrow (0,11)\ [P(11)]$ & 4.1896 & 6046 & $1.51\times10^{-5}$ & $8.36\times10^{-5}$ \\ 
$(1, 11) \rightarrow (0,12)\ [P(12)]$ & 4.2545 & 6461 & $9.76\times10^{-6}$ & $6.24\times10^{-5}$ \\
$(1, 12) \rightarrow (0,13)\ [P(13)]$ & 4.3224 & 6912 & $5.82\times10^{-6}$ & $4.91\times10^{-5}$ \\
$(1, 13) \rightarrow (0,14)\ [P(14)]$ & 4.3936 & 7398 & $3.05\times10^{-6}$ & $3.86\times10^{-5}$ \\
$(1, 14) \rightarrow (0,15)\ [P(15)]$ & 4.4681 & 7918 & $1.20\times10^{-6}$ & $3.15\times10^{-5}$ \\
$(1, 15) \rightarrow (0,16)\ [P(16)]$ & 4.5462 & 8472 & $1.88\times10^{-10}$ & $2.73\times10^{-5}$ \\  
\hline
\multicolumn{5}{c}{$\bm{\upsilon=2\rightarrow1}$} \\
\hline
$(2, 6) \rightarrow (1,7)\ [P(7)]$ & 4.1384 & 8495 & $4.45\times10^{-6}$ & $3.35\times10^{-5}$ \\
$(2, 7) \rightarrow (1,8)\ [P(8)]$ & 4.1952 & 8754 & $4.17\times10^{-7}$ & $3.06\times10^{-5}$ \\  
$(2, 8) \rightarrow (1,9)\ [P(9)]$ & 4.2549 & 9048 & $2.44\times10^{-8}$ & $2.94\times10^{-5}$ \\
$(2, 9) \rightarrow (1,10)\ [P(10)]$ & 4.3175 & 9378 & $3.25\times10^{-9}$ & $2.66\times10^{-5}$ \\ 
$(2, 10) \rightarrow (1,11)\ [P(11)]$ & 4.3833 & 9743 & $9.50\times10^{-10}$ & $2.26\times10^{-5}$ \\ 
\hline
\hline
\end{tabular}
\end{table*}

\end{appendix}

\end{document}